\documentclass{aa}
	\usepackage[english]{babel}
	\usepackage{inputenc}
	\usepackage{txfonts}
	\usepackage{graphics,subfig,graphicx,epstopdf,float,lscape,longtable,dcolumn,color,url,hyperref}

	\usepackage{natbib}
		\bibpunct{(}{)}{;}{a}{}{,} 
	\pdfoutput=1

\renewcommand\thefootnote{\fnsymbol{footnote}} 
	\title{Characterization of Infrared Dark Clouds\footnotemark[1]}
	\subtitle{NH$_3$ Observations of an Absorption-contrast Selected IRDC Sample}
	\author{R.-A.~Chira\inst{\ref{mpia}} \and H.~Beuther\inst{\ref{mpia}} \and H.~Linz\inst{\ref{mpia}} \and F.~Schuller\inst{\ref{eso}} \and C.~M.~Walmsley\inst{\ref{arcetri},\ref{dias}} \and K.~M.~Menten\inst{\ref{mpifr}} \and L.~Bronfman\inst{\ref{uchile}}}
	\institute{Max-Planck Institute for Astronomy, MPIA, K\"onigstuhl 17, 69117 Heidelberg, Germany\\ \email{rox.chira@gmail.com}\label{mpia}
		\and European Southern Observatory, Alonso de Cordova 3107, Casilla 19001, Santiago 19, Chile\label{eso}
		\and Osservatorio Astrofisico di Arcetri, Largo E. Fermi 5, I-50125 Firenze, Italy\label{arcetri}
		\and Dublin Institute for Advanced Studies (DIAS), 31 Fitzwilliam Place, Dublin 2, Ireland\label{dias}
		\and Max-Planck-Institute for Radiostronomy, Auf dem H\"ugel 69, D-53121 Bonn, Germany\label{mpifr}
		\and Departamento de Astronomia, Universidad de Chile, Casilla 36-D, Santiago, Chile\label{uchile}
	}

	\date{Submitted 09.05.2012, Accepted 18.12.2012}

	\abstract{Despite increasing research in massive star formation, little is known about its earliest stages. Infrared Dark Clouds (IRDCs) are cold, dense and massive enough to harbour the sites of future high-mass star formation. But up to now, mainly small samples have been observed and analysed.}
	{To understand the physical conditions during the early stages of high-mass star formation, it is necessary to learn more about the physical conditions and stability in relatively unevolved IRDCs. Thus, for characterising IRDCs studies of large samples are needed.}
	{We investigate a complete sample of 218 northern hemisphere high-contrast IRDCs using the ammonia (1,1)- and (2,2)-inversion transitions. }
	{We detected ammonia (1,1)-inversion transition lines in 109 of our IRDC candidates. Using the data we were able to study the physical conditions within the star-forming regions statistically. We compared them with the conditions in more evolved regions which have been observed in the same fashion as our sample sources. Our results show that IRDCs have, on average, rotation temperatures of 15 K, are turbulent (with line width FWHMs around \linebreak 2 km s$^{-1}$), have ammonia column densities on the order of $10^{14}$ cm$^{-2}$ and molecular hydrogen column densities on the order of \linebreak $10^{22}$ cm$^{-2}$. Their virial masses are between 100 and a few 1000 M$_\odot$. The comparison of bulk kinetic and potential energies indicate that the sources are close to virial equilibrium. }
	{IRDCs are on average cooler and less turbulent than a comparison sample of high-mass protostellar objects, and have lower ammonia column densities. Virial parameters indicate that the majority of IRDCs are currently stable, but are expected to collapse in the future.}

\begin{document}
	\maketitle

\footnotetext{$^{\star}$ Full derivations in Appendix \ref{app_derive} and Figures in \ref{App_bigfig} are published online only. Full Tables \ref{tab01_fit}-\ref{tab04_list_notdet} are only available at the CDS via anonymous ftp to \texttt{cdsarc.u-strasbg.fr (130.79.128.5)} or via
\textcolor{blue}{\url{http://cdsarc.u-strasbg.fr/viz-bin/qcat?J/A+A/}}}

\renewcommand{\thefootnote}{\arabic{footnote}}
	\section{Introduction}\label{intro}
		Despite their relatively small numbers, massive stars are important components of galaxies. However, we know little about the initial conditions for the formation of high-mass stars due to the difficulty of identifying the sites at which they form. Early stages of massive star formation should contain cold gas and dust cores with a peak in the spectral energy distribution around 200 $\mu$m and no emission at near- and mid-infrared wavelengths. Using near- to mid-infrared Galactic Plane surveys from the Midcource Space Experiment (MSX), the Infrared Space Observatory (ISO) and the Spitzer Space Observatory, it is possible to identify large samples of Infrared Dark Clouds (IRDCs). These are dark, compact silhouettes seen against strong Galactic Background emission \citep{Simon2006,Peretto2009,Perault1996,Egan1998}. IRDCs have high molecular hydrogen column densities \citep[$\sim 10^{22}$ cm$^{-2}$,][]{Simon2006} and block background radiation, in particular at optical and near-infrared wavelengths. They 
cover broad ranges of gas masses from low-mass to high-mass star-forming regions. The IRDCs on the upper end of gas mass range are considered to be capable of harbouring high-mass star-forming regions in their earliest evolutionary stages. Based on the MSX database, \citet{Simon2006} built up a catalogue of about \linebreak 10,000 candidate IRDCs which are excellent starting points for studies of the early stages of high-mass star formation. \\
		Since this catalogue is only based on mid-infrared absorption images, it lacks any additional information about the physical conditions within these regions. Therefore, further studies are needed. In this work we investigate the temperatures, turbulent properties and the virial parameters of IRDCs using the ammonia (NH$_3$) $(J,K) =$ (1,1)- and (2,2)-inversion transitions. Ammonia has critical densities on the order of \linebreak 10$^3$ cm$^{-3}$ and so its radio wavelength inversion transitions are good higher-density tracers which suffer from minimal freeze out \citep{Bergin2007}. The (1,1)- and (2,2)-inversion lines can be observed simultaneously making ammonia also a good thermometer \citep{Walmsley1983}. Since radiative transitions between different K-ladders are forbidden, metastable (J,K)-inversion level (with J = K) are only populated via collisions \citep{Ho1983}. From the observed ammonia spectra it is possible to determine a source's kinetic temperatures \citep{Walmsley1983}, the level of 
turbulence via the line widths and its virial mass. Using the lines' rest velocities we can also calculate the kinematic distances of the sources using a model of the Galactic rotation curve \citep{Reid2009}. \\
		Ammonia studies analysing star-forming gas clumps have also been conducted for less massive sources \citep[e.g.][]{Harju1991,Harju1993}. It has been observed in high-mass regions for decades \citep{Churchwell1990,Molinari1996,Sridharan2002,Schreyer1996}, but most previous studies discussed smaller and/or more evolved samples \citep[e.g.,][]{Sridharan2005,Pillai2006}. The advantage of our sample is that it is large enough to study the physical properties of IRDCs statistically. Thus, we can compare our results to studies of objects in more evolved stages of high-mass star formation. In particular, we compare the physical conditions found in IRDCs with those in High-Mass Protostellar Objects \citep[HMPOs, by][]{Sridharan2002}. HMPOs comprise the next evolutionary stage where at the core centres massive protostars have formed already. Typical luminosities exceed are $10^4\,\mbox{L}_\odot$ and accretion processes are likely still ongoing \citep{Sridharan2002}. \\
		Furthermore, the ammonia data will be compared with data on submm wavelength ($870~\mu$m) dust emission collected in the course of the ATLASGAL (the Atacama Pathfinder Experiment Telescope Large Area Survey of the Galaxy) survey \citep{Schuller2009}. With this additional data, we will be able to determine the sources gas masses, molecular hydrogen column densities and ammonia abundances and gain information about the stability of IRDCs. \\
		Recently, \citet{Dunham2011} and \citet{Wienen2012} presented their works. They studied more heterogeneous samples (concerning the evolutionary stages of star formation in the observed regions) based on BOLOCAM and ATLASGAL \citep{Rosolowsky2010,Schuller2009}. Although they did not focus on IRDCs as we did, they have corresponding subsamples with sources at typical distances similar to ours. Thus, we will discuss similarities and differences between these studies and ours as well. \\
		Further details about the selection criteria of our IRDC sample, the observations, data reduction and analysis are given in the Sects. \ref{sample} - \ref{results}. In Sect. \ref{results}, the IRDCs' rotation and kinetic temperatures, and column densities of ammonia are derived and compared with the data of the HMPO sample by \citet{Sridharan2002}. Furthermore, we study the molecular hydrogen column density and ammonia abundance, as well as the stability of the IRDCs by analysing their virial parameters. \\
		\vspace*{0.5cm}
		The paper is outlined as follows. Sect. \ref{sample} introduces details about our sample and the selection criteria. Sect. \ref{observation} provides information about the observations and data reduction. In Sect. \ref{results} we present and discuss our results. In Sect. \ref{summary} we summarise them and state our conclusions.

	\section{Source Sample}\label{sample}
		Based on the 8.3 $\mu$m data collected by the MSX, \citet{Simon2006} compiled a catalogue of more than 10,000 candidate IRDCs. Those contain about 12,000 embedded Infrared Dark Cores that appear as high-contrast peaks within the IRDCs. We specifically selected Cores and refer to them here as IRDCs. In an attempt to present a complete sample of the northern hemisphere high-contrast IRDCs, our sample of 218 sources was selected from the catalogue of \citet{Simon2006} based on the following criteria. \\
		Firstly, the sources had to have contrast values greater than 0.3, where contrast is defined as \linebreak contrast = (background $-$ image) / background \citep[see][]{Simon2006}. This selects clouds with column densities of the order of \linebreak $10^{22}$ cm$^{-2}$ of molecular hydrogen or $10^{14}$ cm$^{-2}$ of ammonia. For a source to be considered significant, the ratio of the peak-contrast versus its uncertainty has to be larger than three. \\
		Secondly, the sources needed to be observable from Effelsberg for more than five hours and be larger than \linebreak 0.6 arcmin$^2$ to fill the beam of $40''$ FWHM at the given frequencies. Therefore, they have to be above an elevation of 20$^\circ$ corresponding to Galactic Longitudes roughly larger than 19$^\circ$. Thus, our systematic analysis of our sample started at $l = 19.27^\circ$ (it was also possible to observe nine sources with $l < 19^\circ$ which are exceptions). The coordinates of the IRDCs we used, were taken from the catalogue based on the MSX observations \citep{Simon2006}. We see small differences compared to studies using (sub)mm peak positions. For consistency, we used the same positions for the ATLASGAL and Spitzer maps (cf. Sect. \ref{virial_gas} and Sect. \ref{ammonia_abu}). 

	\section{Observation and Data}\label{observation}
		In 2008 and 2009, we used the Effelsberg 100\,m telescope to observe simultaneously the NH$_3$ (1,1)- and (2,2)- inversion lines with central rest frequencies of 23.6945 and 23.7226\,GHz, respectively. The data were taken in a frequency switching mode with on average five minutes integration time per position. The weather was adequate with system temperatures mostly below 50 K. The full width half maximum (FWHM) is 40$''$ and the velocity resolution is $\Delta v=0.38$\,km s$^{-1}$. The absolute flux calibration was determined from observations of NGC 7027. \\
		Fitting the whole hyperfine structure (HFS) of both lines using the standard procedures of CLASS \citep{gildas_class}, the brightness temperatures, $T_{\rm{mb}}$, local standard of rest velocities, $V_{\rm{lsr}}$, line width FWHMs, $\Delta$v, and optical depths, $\tau$, were derived. The results are listed in Table \ref{tab01_fit}. \\ 
		One has to consider that the HFS satellites, especially of the (2,2) transition, have not been detected in every spectrum. This will be discussed in Sect. \ref{detec_rate}. 

		\begin{table}[H]
			\begin{minipage}{0.5\textwidth}
				\caption{Number of sources being eliminated by the individual selection criteria.}
				\centering
				\begin{tabular}[]{l|c}
					\hline
					Selection Criterion & Number of eliminated Sources\footnote{Note that some sources have been eliminated, because they did not fulfil several of these criteria. Therefore, the sum of the listed sources here is bigger than the 109 sources which have been actually eliminated.} \\ \hline
					(a) NH$_3$ (1,1) line detected & 91 \\
					(b) 8 $< \mbox{T}_{\mbox{rot}} <$ 50 K & 17 \\
					(c) F$_{870 \mu{\rm m}}^{40''} > $ 4 RMS & 8 \\ \hline
				\end{tabular}
				\label{kickout_stat}
			\end{minipage}
		\end{table}

		\begin{figure}[ht]
			\centering
			\subfloat[(1,1)-inversion line]{
			\includegraphics[angle=270,width=0.45\textwidth]{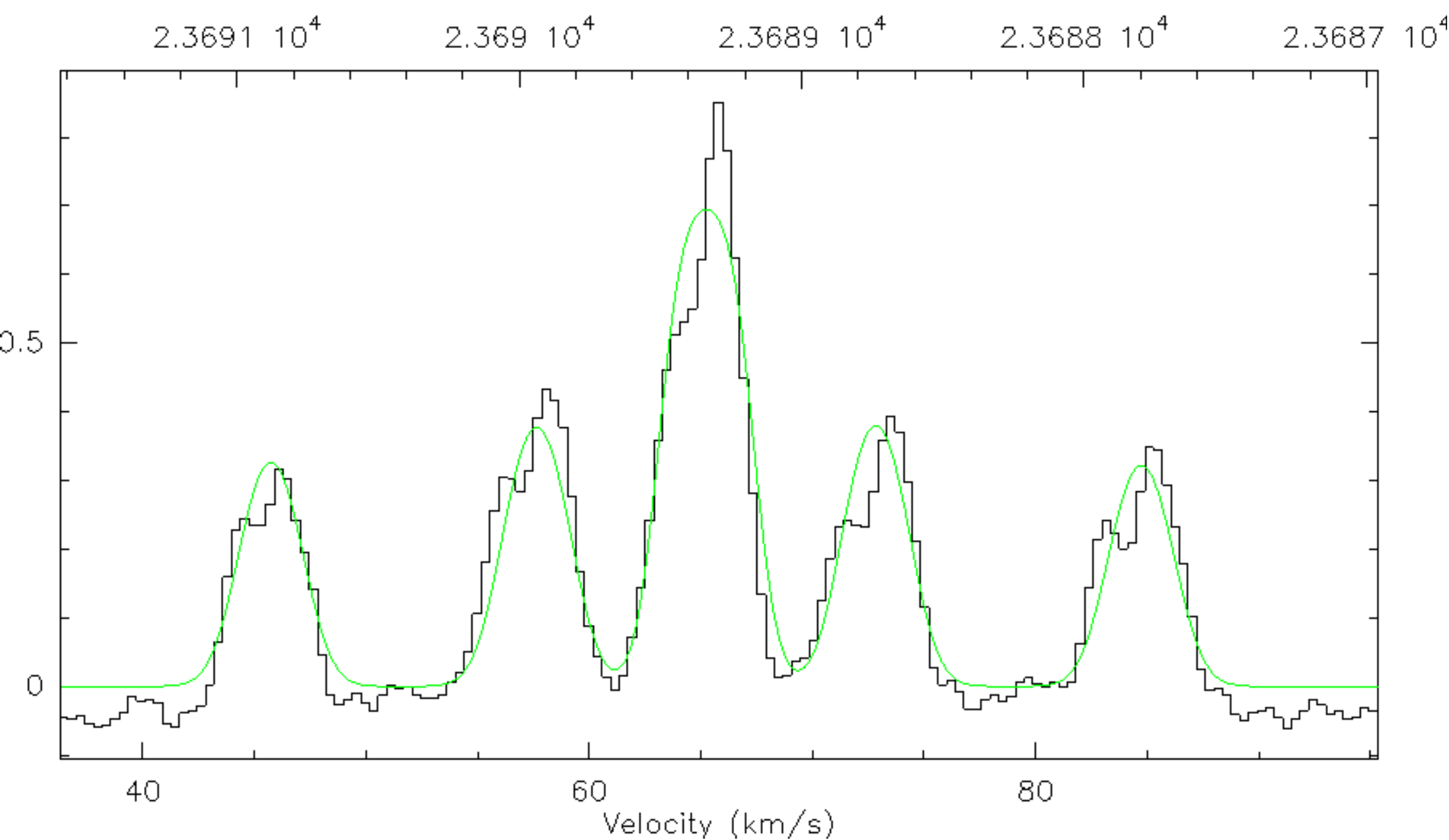}
				\label{9901_good_11}
			} \\
			\subfloat[(2,2)-inversion line]{
				\includegraphics[angle=270,width=0.45\textwidth]{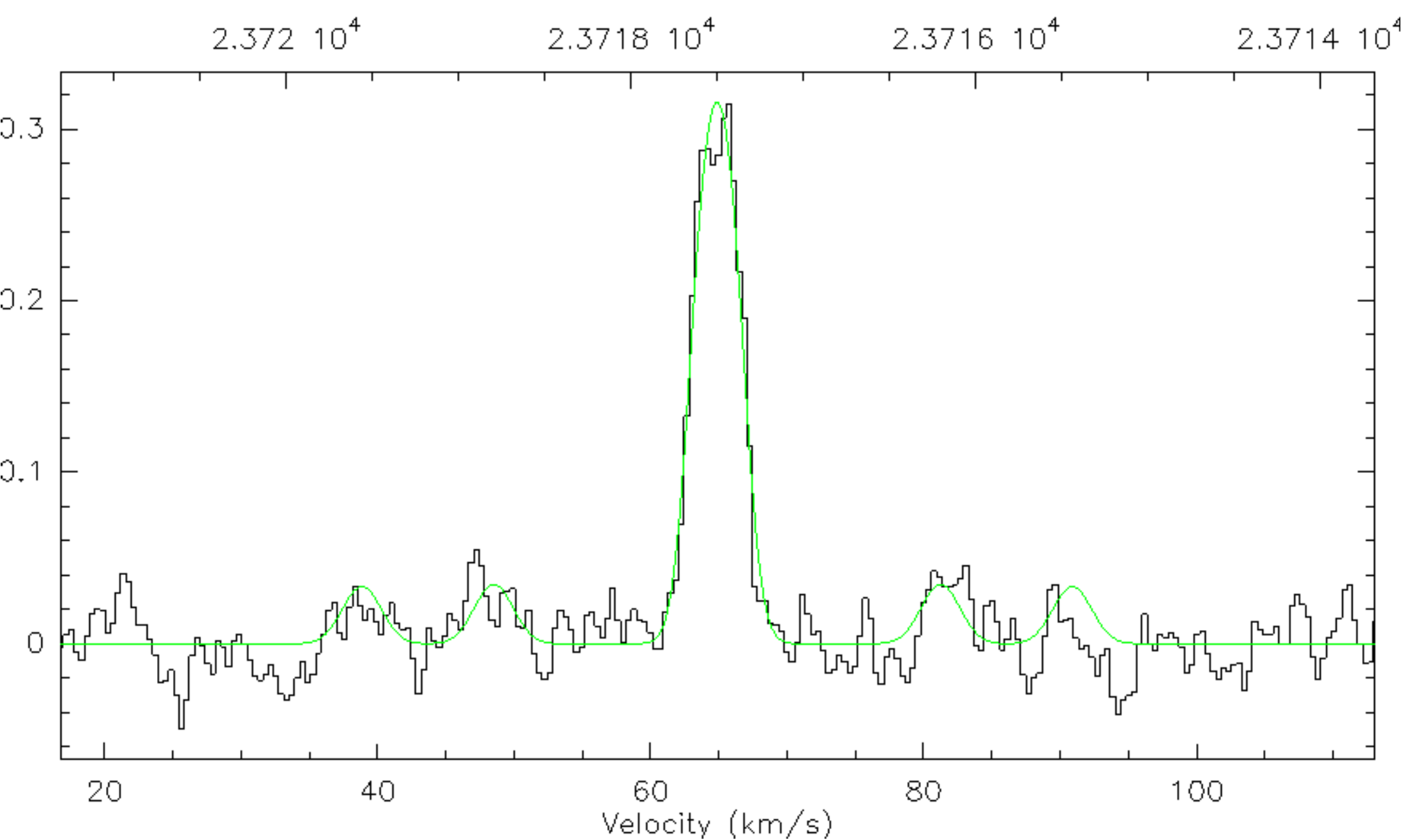}
				\label{9901_good_22}
			}
			\caption{Example for a good detection in ammonia. NH$_3$ inversion lines in G19.92-00.29A. The green line indicates the hyperfine fit by CLASS.}
			\label{9901_good}
		\end{figure}

	\section{Results}\label{results}
		\subsection{Distribution of Detected IRDCs within the Galactic Plane}\label{detec}
			\subsubsection{Detection Rate}\label{detec_rate}
			Out of our original sample of 218 sources 109 were detected in ammonia. The fit parameters of the sources with detected ammonia inversion transitions are listed in Tables \ref{tab01_fit} to \ref{tab03_gas} in the Appendix. The formal errors (1$\sigma$) for the calculated quantities have been derived from Gaussian error propagation. IRDCs positions are indicated in Figs. \ref{App_fig01} and \ref{App_fig02} (cf. Appendix). The black-and-white background is taken from the ATLASGAL dust emission maps by \citet{Schuller2009}. The figures show a relatively good agreement between the compact sources detected by MSX extinction and by ATLASGAL. \\
			For our analysis we applied the following criteria to obtain the significance of the statistics. \\
			(a) The (1,1)-inversion line is detected and the brightness temperatures are higher than the threefold root-mean-square (RMS). \\
			(b) The rotation temperatures are between 8 K and 50 K and the absolute errors are lower than 20 K. This is necessary, because both inversion lines are sensible temperature indicators, but only within this range \citep{Danby1988}. \\
			(c) The flux densities F$_{\nu}^{40''}$ of the IRDCs' counterparts in the ATLASGAL dust emission maps are greater than four times the RMS noise. This last criterion is needed for the virial analysis (cf. Sect. \ref{virial}). In this case we were not able to derive independent gas masses or hydrogen column densities. If a source fulfils the first two criteria, but not the last one, its NH$_3$ properties - rotation and kinetic temperatures, column densities of ammonia and the virial masses \citep{Schilke,Ho1983,MacLaren1988} - have been calculated, anyway, but not the virial parameter, hydrogen column density and ammonia abundance. \\
			A good example for a detection fulfilling all of these criteria is G19.92-00.29A which is shown in Fig. \ref{9901_good}. The green lines represent the fits to the HFS lines determined by CLASS. \\
			81 out of 218 sources satisfied all our criteria. There were 29 sources without detectable (2,2)-inversion lines although the (1,1)-inversion lines was detected. In this case, it is not possible to derive exact parameters. But it is possible to give upper limits for the rotation temperature by using three-times the average RMS noise (0.0925 K, cf. Sect. \ref{app_temp}) as an upper limit for the (2,2)-line intensity \citep[following][cf. Sect. \ref{rot_temp}]{Ho1983}.
			In 91 sources the ammonia inversion transitions are not detected. For the sources not satisfying criterion (a) it is not possible to do any further calculations and analysis. There are 25 spectra with (1,1)-inversion lines lying within the noise. In this case the hyperfine fits by CLASS were not of good quality. A list of all sources toward which ammonia could not be detected is given in Table \ref{tab04_list_notdet}.  \\
			17 sources were dropped, because their calculated rotation temperatures were not between 8 and 50 K. This could be caused by poor signal-to-noise ratios and large errors in the hyperfine structure fits (cf. criterion (a)). \\
			Eight sources turned out to have no significant counterparts in the ATLASGAL dust emission maps. As explained before we used the ammonia data to calculate temperatures, line widths and ammonia column densities for these sources and we list them in the tables in Appendix \ref{app_data}. \\
			Our sample and that of \citet{Wienen2012} have very little overlap. Depending on the search radii (defining the area around our sources where we searched corresponding sources of \citealt{Wienen2012}; in our case $20''$, $40''$, or $60''$) we find 10, 16, or 22 sources in common. The properties derived by \citet{Wienen2012} match our results for the corresponding sources. \\
			Our NH$_3$ detection rate is 50\%. This lower than rates of 87\% and 72\% (respectively from \citealt{Wienen2012} and \citealt{Dunham2011}) for samples which based on (sub)mm emission data. This indicates that sample selection solely based on mid-infrared absorption studies against the Galactic background are less reliable. Even if a contrast criterion is enforced, such studies can pick up a variety of structures that emulate IRDCs but might often turn out to be blown-out bubbles with very little dense gas inside \citep[e.g.][]{Wilcock2012b}. Comprehensive (sub)mm emission studies were not available at the time of our sample selection. The best future sample selections will be most likely based on a combination of mid-infrared absorption and (sub)mm emission selection criteria.

			\begin{figure}[htb]
				\centering
				\includegraphics[width=0.5\textwidth]{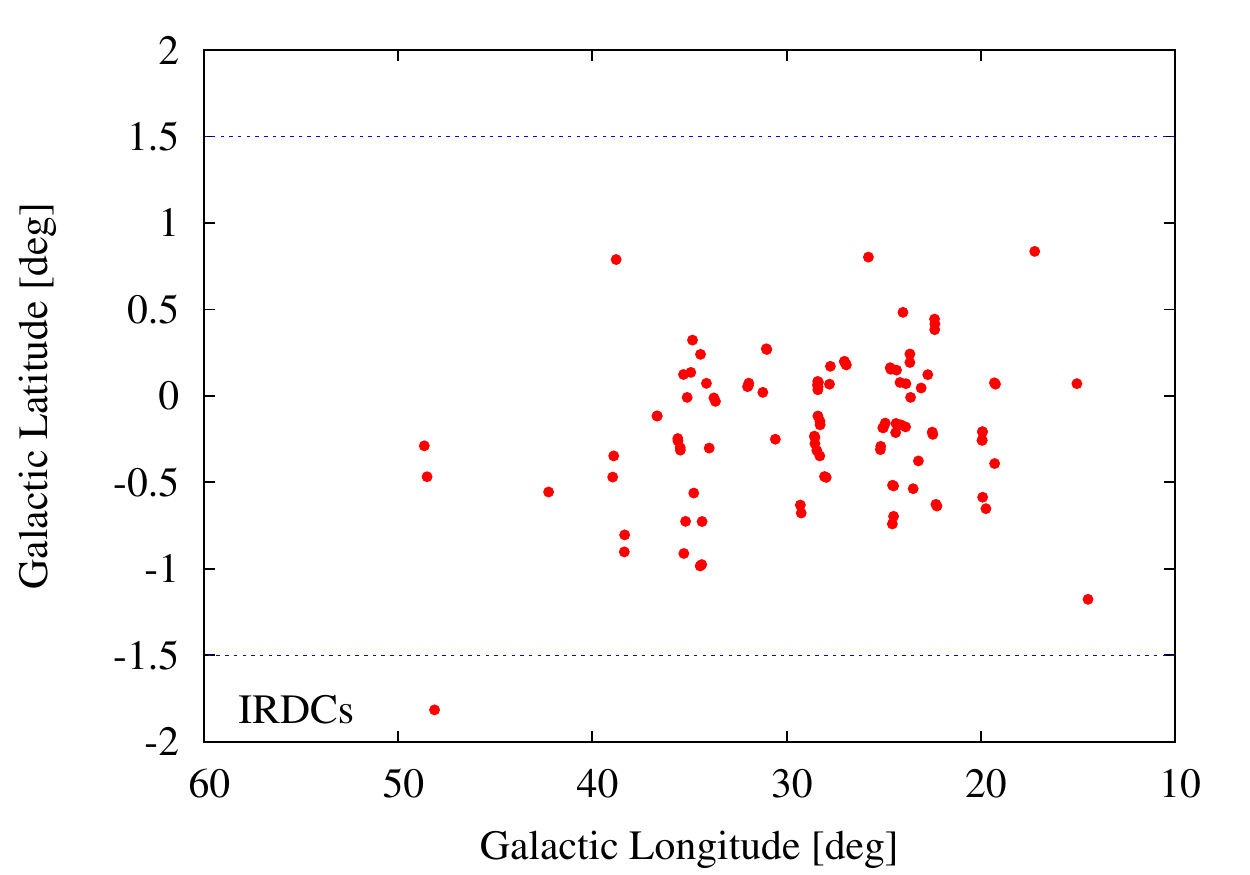}
					\caption{Positions of selected IRDCs with detected ammonia transitions in the Galactic Plane.}
					\label{pos_plane}
				\includegraphics[width=0.5\textwidth]{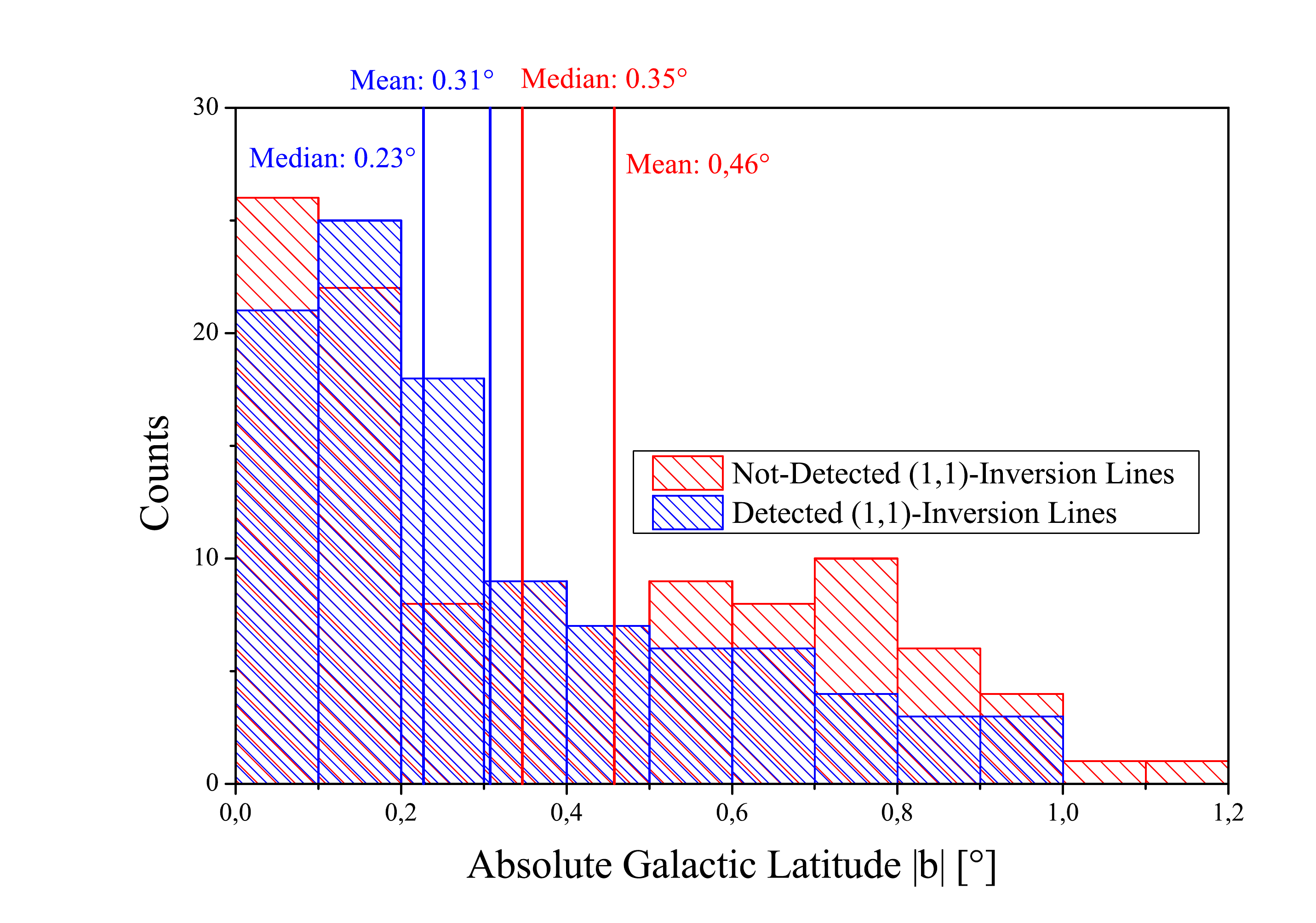}
					\caption{Distribution of IRDCs with detected and without detected ammonia (1,1)-inversion line along the absolute Galactic Latitude.}
					\label{det_stat}
			\end{figure}
			\begin{figure}[hbt]
				\centering
				\includegraphics[width=0.5\textwidth]{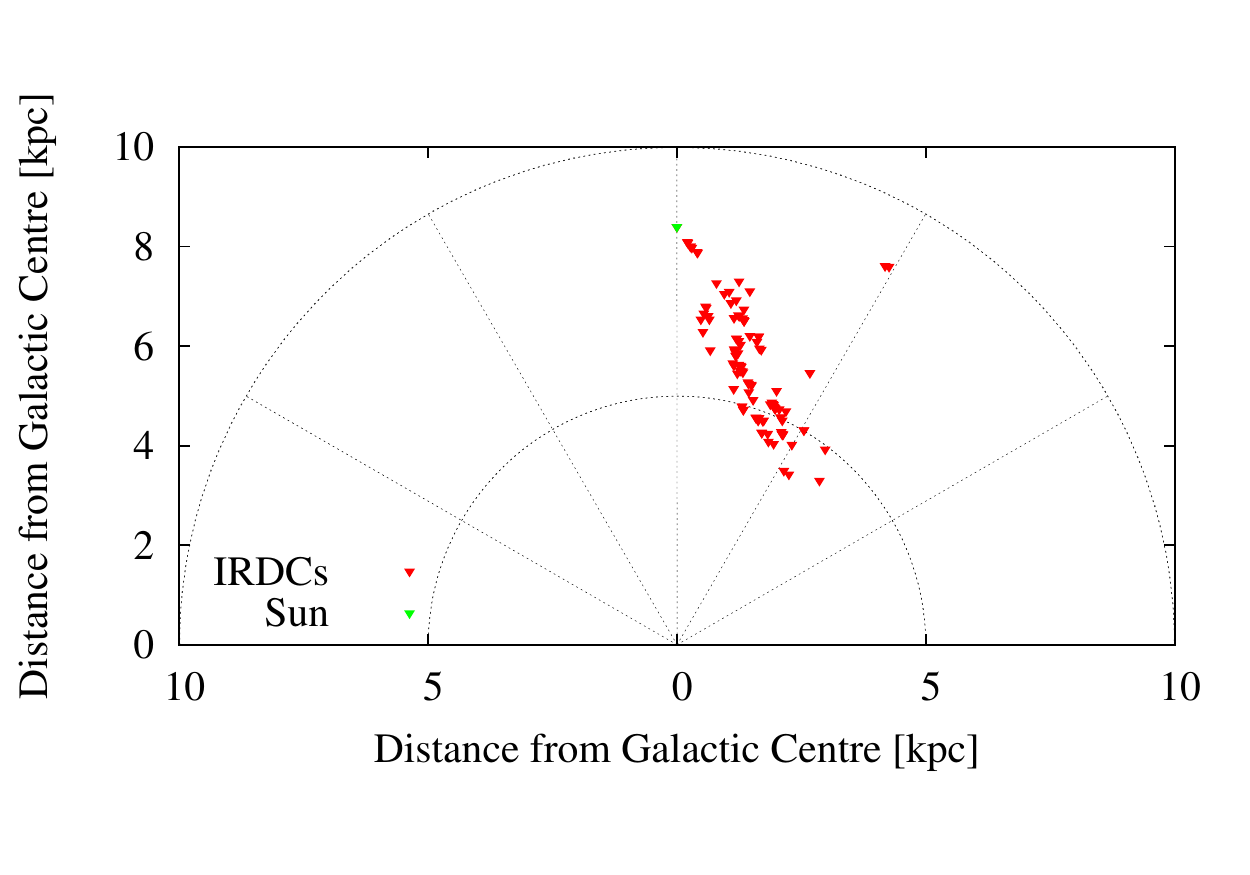}
					\caption{Positions of selected IRDCs within the Galaxy.}
					\label{pos_galaxy}
			\end{figure}

			\subsubsection{Distribution within the Galactic Plane}\label{detec_distri}
			\noindent Fig. \ref{pos_plane} shows the positions of the selected sources within the Galactic Plane. The IRDCs lie close to the Galactic Midplane (Gal. Latitude $b \sim$ 0$^\circ$) where the gas density tend to be higher \citep{Schuller2009,Rosolowsky2010,Beuther2012}. As described in \citet{Beuther2012} in more detail the distribution along the Galactic Latitude is symmetric and its peak is close to the Galactic Mid-Plane ($\sim 0.1^\circ$). Therefore observations within a latitude range of $|b|\,\leq$ 0.5$^\circ$ are recommended to reliably identify IRDCs. \\
			\noindent In Fig. \ref{det_stat} the distributions of the sources with detected and non-detected ammonia (1,1)-inversion line have been plotted against the absolute value of their Galactic Latitude b. For both distributions the mean values and medians have been calculated and are marked in Fig. \ref{det_stat}. The IRDCs with detected inversion lines (mean: (0.32$\pm$0.02)$^\circ$, median: 0.24$^\circ$) are significantly closer to the Galactic Mid-Plane than those without (mean: (0.47$\pm$0.05)$^\circ$, median: 0.35$^\circ$). This implies that identification of IRDCs based on absorption studies against the Galactic background are less reliable at larger Galactic Latitudes $b$. \\
			\citet{Schuller2009} and \citet{Beuther2012} have found a symmetric distribution of compact ATLASGAL sources around the mid-plane of the Milky Way. This is similar to our distribution of the sources where the ammonia (1,1)-inversion transition could be detected. But they have also detected a shift of the distribution peak to $b \sim -0.1^\circ$. However our sample is too small to do a similar analysis.

			\subsubsection{Distance}\label{dist}
			In this and the following sections, the results for the physical conditions and virial parameters will be presented and discussed. Therefore, we only take the 109 sources into account which were detected in the ammonia.  \\
			To be able to calculate the IRDCs' virial masses, which will be averaged over the beam, we needed the sources' radii. For estimating them, we derived the kinematic distances between the sources and the sun. Using a model for the Galactic rotation with new parameters proposed by \citet{Reid2009}, (near and far) kinematic distances of a source with measured LSR velocity at given Galactic coordinates can be derived. The method is described in more details in the paper by \citet{Reid2009}. Given that the sources in our sample appear in silhouette against background radiation, we chose the near values for sources with a distance ambiguity. \\
			The IRDCs are at distances of a few kilo-parsecs (cf. Table \ref{IRDC_stat} and Table \ref{tab02_initial}) and have radii between 0.25 pc and 0.5 pc. To estimate these, we assumed the FWHM of the Effelsberg beam (40'') as a proxy for the size. Their positions within the Galactic Plane and Galaxy are sketched in the Figs. \ref{pos_plane} and \ref{pos_galaxy}.

		\subsection{Rotation and Kinetic Temperature}\label{rot_temp}
			For deriving the rotation temperatures and column densities, we followed the outlines by \citet{Schilke}, \citet{Ho1983} and \citet{Ragan2011}. For a better understanding we shortly summarise the most important points in Appendix \ref{app_derive}. \\
			The calculated results for all sources are presented in Table \ref{tab02_initial} in the Appendix. The errors have been derived from Gaussian error propagation. The same applies to all other quantities if not mentioned otherwise. \\ 
			\noindent The statistical properties of the sample are summarised in \linebreak Table \ref{IRDC_stat} including the mean values and the medians of the derived quantities, as well as the minimal and maximal values. \\
			\begin{figure}[tb]
				\centering
				\includegraphics[width=0.5\textwidth]{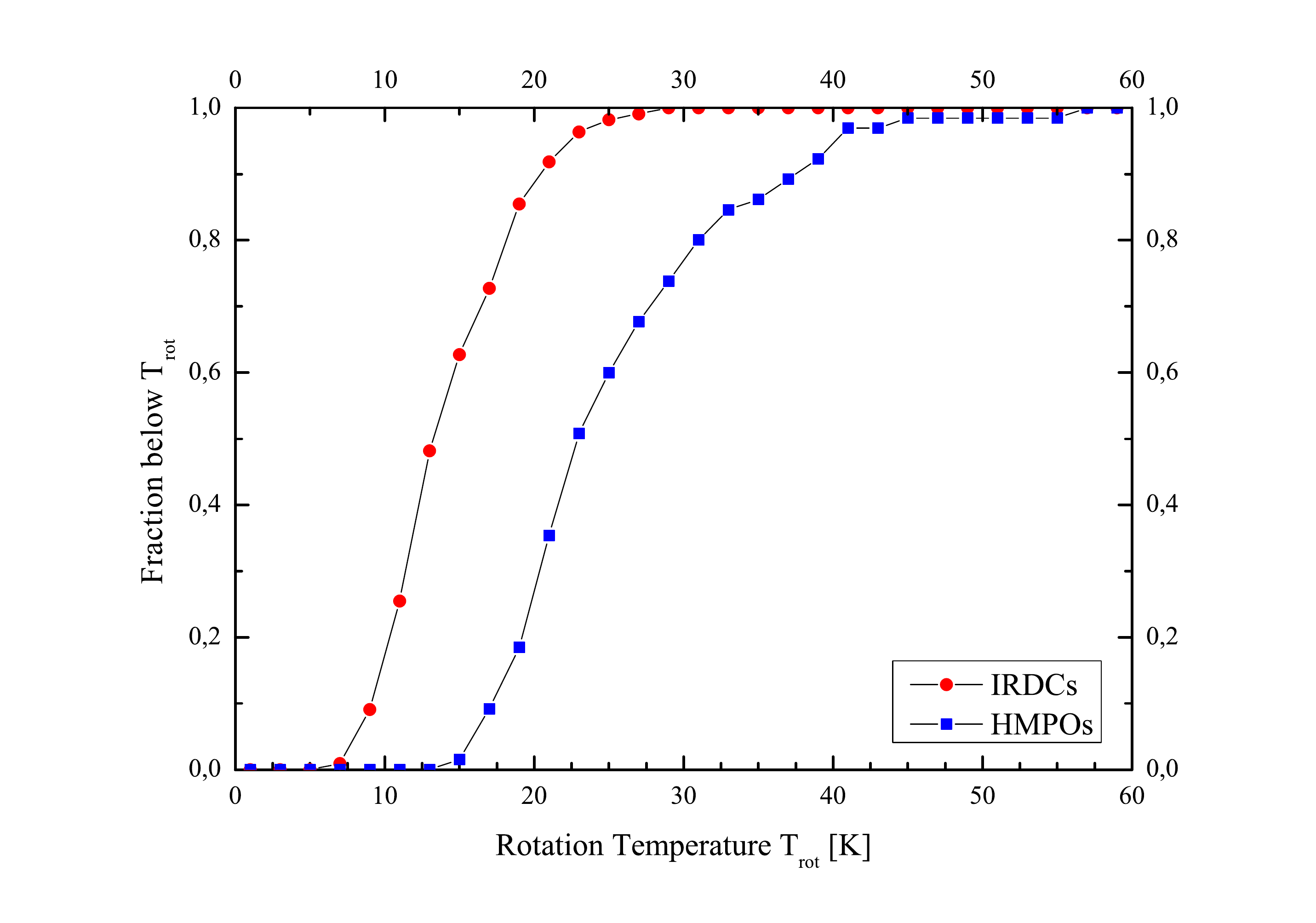}
					\caption{Cumulative distributions of NH$_3$ rotation temperatures of IRDCs (red data points) and HMPOs (blue data points).}
					\label{frac_Trot} 
 				\includegraphics[width=0.5\textwidth]{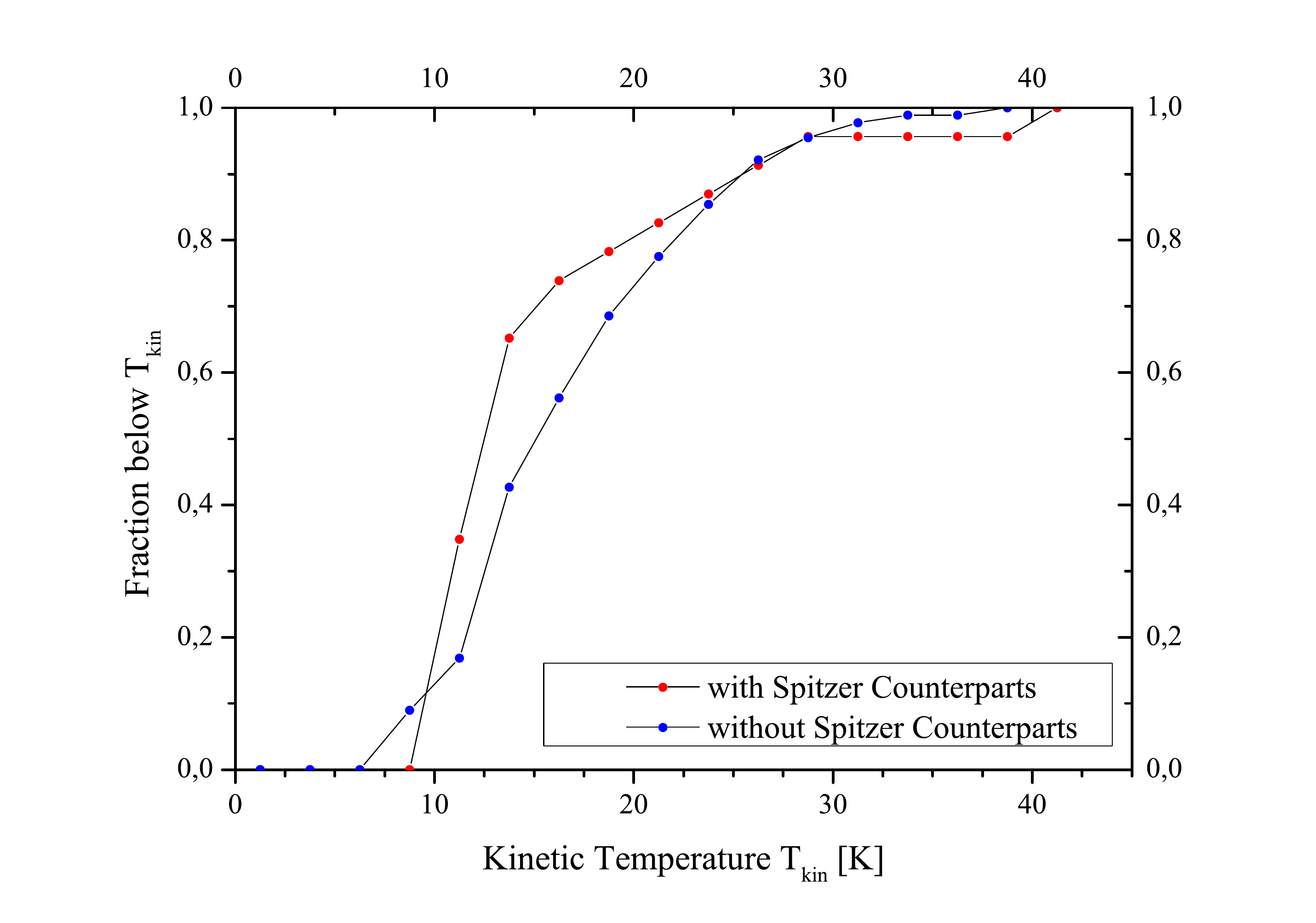}
 					\caption{Cumulative distributions based on the sources where NH$_3$ has been detected. The red subsample includes the sources with Spitzer 24 $\mu$m association, the blue subsample the ones without. }
 					\label{Tkin_spitzer}
			\end{figure}
			\noindent The IRDCs have rotation temperatures between 8 K and \linebreak 29 K. Although the latter being an upper limit, the majority of the IRDCs have mean rotation temperatures around 15 K (median: 14 K) being more obvious in Fig. \ref{frac_Trot}. It shows a cumulative distribution of our sample's rotation temperatures (red data points). About 80\% of the sources have values between 10 and 20 K. This is consistent with previous studies \citep{Pillai2006,Peretto2010,Ragan2011}. \\
			In addition, the blue points show the cumulative distribution of rotation temperatures derived for the sample of 65 high-Mass Protostellar Objects (HMPOs) from \citet{Sridharan2002}. This sample had been observed in the same fashion as ours with the Effelsberg telescope. Therefore, the results for rotation temperatures, line width FWHMs and column densities can be compared with the results for our sample IRDCs here. HMPOs are considered to be the next evolutionary step in high-mass star formation after IRDCs \citep{Beuther2007}.  \\
			The sample of HMPOs has a mean rotation temperature of \linebreak 26 K (median: 24 K) which is significantly higher than the value we find for our IRDCs. Therefore, the rotation temperature is a good indicator for differentiating between evolutionary stages: the rotation temperature increases with evolution. While HMPO temperatures and luminosities mainly stem from internal heating processes (hydrogen burning and accretion), the IRDC temperatures and luminosities are dominated by external radiation fields \citep{Wilcock2012b} . \\
			\noindent We derived kinetic temperatures with the approximation of \citet{Tafalla2004}. More details are given in Appendix \ref{app_temp}. We estimate that they have errors of 5\%. The kinetic temperatures of the IRDCs vary between 8 K and 41 K with a mean value of 18 K and a median of 16 K. These values are similar to those of the rotation temperature. This is not surprising, because both quantities are nearly linearly connected until \linebreak 20 K and, thus, approximately equal within this regime. \citep{Danby1988}.
			\noindent Our results match those of \citet[median T$_{\mbox{\tiny rot}}$ of 17 K]{Wienen2012} and \citet[mean T$_{\mbox{\tiny kin}}$ of 15.6 K]{Dunham2011}. Furthermore, the trend toward higher rotation temperatures and line widths (cf. Sect. \ref{width}) with progressing protostellar evolution agrees with the results of \citet{Wienen2012} who compared the relative number and cumulative distributions of different stages between IRDCs and the envelopes of ultracompact HII regions. \\
			Comparing our results with those of \citet{Harju1991,Harju1993} (mean  T$_{\mbox{\tiny kin}}$ between 14 and 20 K), our sources have, on average, the same temperature or are slightly cooler. This can be explained by taking into account that \citeauthor{Harju1991} included also sources being associated with IRAS sources heating up their envelopes. \\
			Using Spitzer 24$\mu$m data we checked for associations with Young Stellar Object (YSO) to identify a sample of starless IRDCs. We looked for any emission source within a circular area of radius 20'' around the position. The results are listed in the last column of Table \ref{tab02_initial}. We found that 87 (79.8\%) IRDC candidates have no counterpart. But 22 (20.2\%) of our sample IRDCs are associated with YSOs or lie close enough to a Spitzer 24$\mu$m source that the beam includes parts of them. We also investigated whether there exists a correlation between the (non)detection in Spitzer 24$\mu$m and the kinetic temperature. However, no statistically significant difference between the two subsamples could be identify (cf. Fig. \ref{Tkin_spitzer}). This could be explained by considering our beam which covers a greater volume than the still too small and too young protostars are able to influence. While the line widths of sources with and without Spitzer counterparts are almost identical, there is a 
trend of higher H$_2$ and NH$_3$ column densities for sources with Spitzer counterparts. However, since the sample with Spitzer 24$\mu$m detections is relatively small (22), it is unclear whether this is statistically significant.

			\begin{table}[th]
				\begin{minipage}[]{0.5\textwidth}
					\centering
					\caption{Statistics of IRDC sample. The statistics have been built up by the 109 IRDCs with detected ammonia transitions.}
					\begin{tabular}[]{l|c|ccc}
						\hline
						IRDC Parameter\footnote[2]{Note that T$_{\mbox{\tiny rot}}$ represents the rotation temperature, T$_{\mbox{\tiny kin}}$ the kinetic temperature, N$_{NH_3}$ the column density of ammonia, N$_{H_2}^{19''}$ the column density of molecular hydrogen, $\chi_{\mbox{\tiny NH$_3$}}$ the abundance of ammonia, $\Delta$v$_1$ the line width FWHM of the ammonia (1,1)-inversion line, d the distance between source and sun, M$_{\mbox{\tiny vir}}$ the virial mass and $\alpha_{\mbox{\tiny i}}$ the virial parameters for the used models (cf. Sect. \ref{virial}).} & Mean & Min. & Median & Max. \\ \hline
						T$_{\mbox{\tiny rot}}$ [K] & 15 & 8 & 14 & 29 \\
						T$_{\mbox{\tiny kin}}$ [K] & 18 & 8 & 16 & 41 \\
						N$_{NH_3}$ [10$^{14}$ cm$^{-2}$] & 7.0 & 0.9 & 5.1 & 32.0 \\
						N$_{H_2}^{19''}$ [10$^{22}$ cm$^{-2}$] & 3.5 & 0.1 & 2.9 & 27.3 \\
						$\chi_{NH_3}$ [10$^{-8}$] & 3.8 & 0.6 & 1.9 & 45.9 \\
						$\Delta$v$_1$ [km s$^{-1}$] & 1.7 & 0.7 & 1.6 & 4.0 \\
						d [kpc] & 3.2 & 0.4 & 3.2 & 5.9 \\ \hline
						M$_{\mbox{\tiny vir}}$\footnote[3]{The here given virial mass M$_{\mbox{\tiny vir}}$ is the mean value of the virial masses for the individual density profiles. Cf. Sect. \ref{virial_mass} for more details.} [M$_\odot$] & 325 & 6 & 264 & 1099 \\
						$\alpha_{\mbox{\tiny OH}}$\footnote[4]{These are the sources' virial parameters, giving the ratio between kinetic and gravitational energy and, thus, an estimation for the stability of the sources. These Parameters will be discussed in more detail in Sect. \ref{virial_parameter}.} & 6.1 & 0.3 & 2.3 & 155.2 \\
						${\alpha_{\mbox{\tiny MRN}}}\footnotemark[4]$ & 1.9 & 0.1 & 0.7 & 49.4 \\ \hline
					\end{tabular}
					\label{IRDC_stat}
				\end{minipage}
			\end{table}

			\begin{figure}[t]
				\centering
				\includegraphics[width=0.5\textwidth]{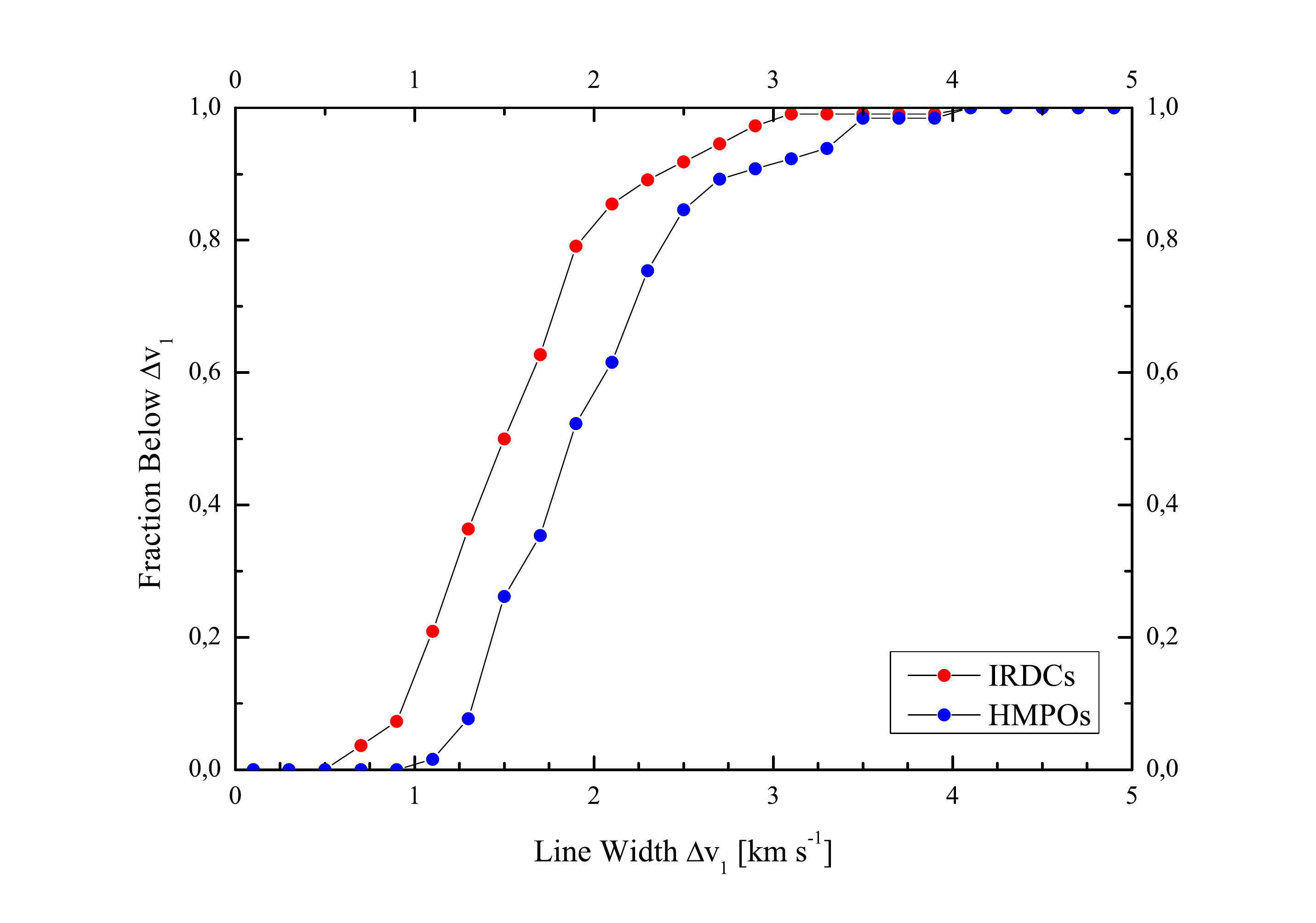}
					\caption{Cumulative distributions of NH$_3$ line width FWHMs of IRDCs (red data points) and HMPOs (blue data points). }
					\label{frac_width}
			\end{figure}

		\subsection{Line Width}\label{width}
			Another observable to characterise IRDCs and HMPOs is the line width. The IRDCs have NH$_3$ (1,1) line widths of around 1.7 km s$^{-1}$ (mean and median) which agrees with the results of \citet{Wienen2012}, but is higher than the values of \citet{Harju1991,Harju1993}. The latter can be explained by taking into account that the sources studied by \citet{Harju1991} are at much smaller distances than ours. Thus, for the same beam one sees smaller size scales and hence smaller FWHMs which agrees with the Larson line width - size relation. \\
			\noindent We investigated possible correlations between the FWHM and the H$_2$ column density as well as between the line width and the kinetic temperature. However, no clear correlation between these quantities was found.  \\
			\noindent Fig. \ref{frac_width} shows that the IRDCs are not only colder than the HMPOs, but have also narrower lines (for HMPOs: mean: 2.1 km s$^{-1}$, median: 2.0 km s$^{-1}$). The line widths are larger than expected for thermal emission at the given temperatures \linebreak ($\approx$ 0.2 km s$^{-1}$) observed in low-mass star-forming regions \citep{Knuth1996}. The increased line widths can be caused by intrinsic cloud turbulence, infall motions or outflows from young forming stars.

		\subsection{Ammonia Column Density}\label{col_dens}
\begin{figure}[t]
				\centering
				\includegraphics[width=0.5\textwidth]{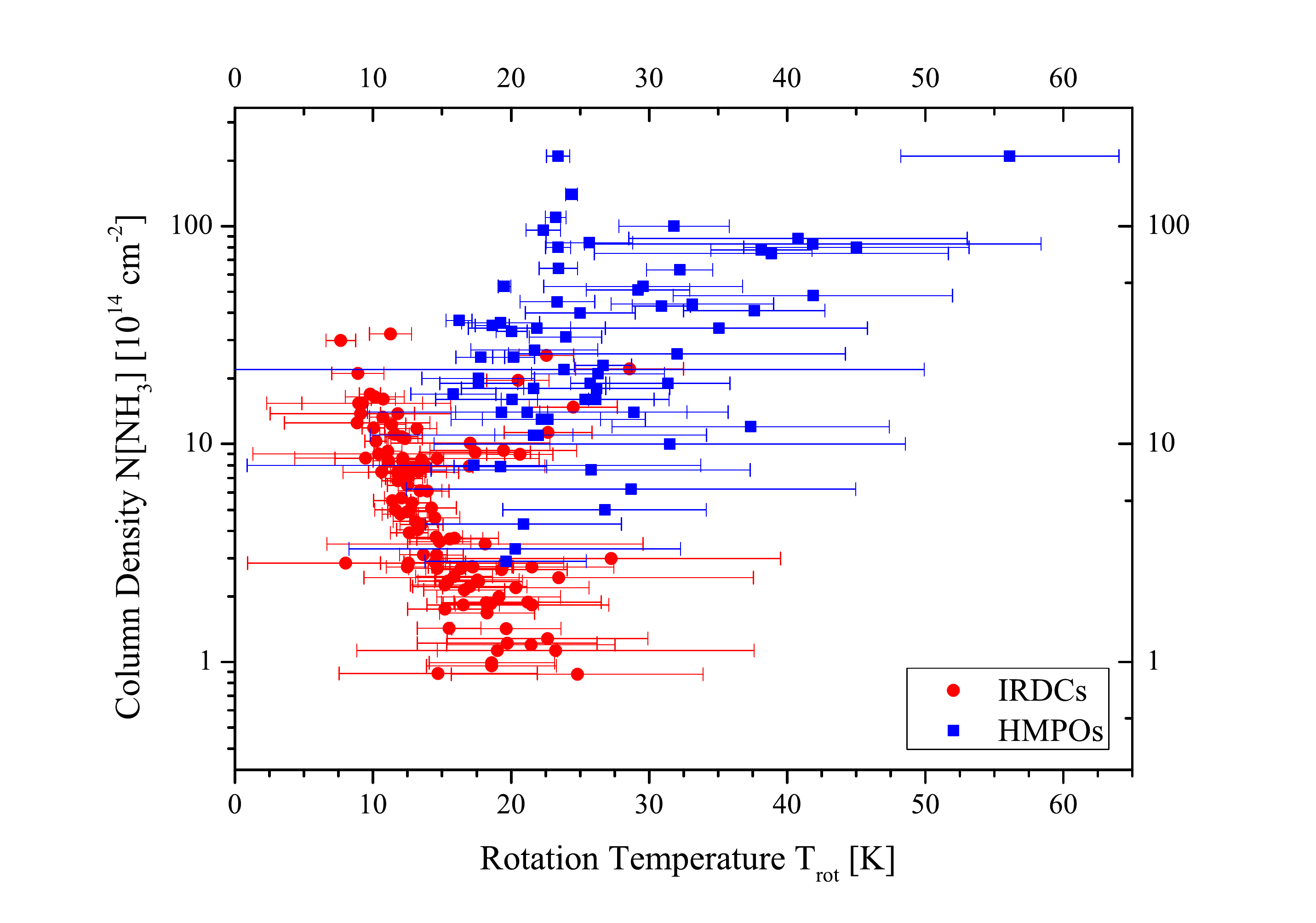}
					\caption{Correlation plot between the NH$_3$ total column density and the NH$_3$ rotation temperature of the IRDCs (red data points) and the HMPOs (blue data points).}
					\label{Trot_density}
			\end{figure}

			Analogous to the temperatures the derivation of the column density can be found in Appendix \ref{app_dens}. The column densities for the IRDCs are listed up in Table \ref{tab02_initial} in the Appendix. \\
			The ammonia column densities range between \linebreak 9$\cdot 10^{13}$ cm$^{-2}$ and 3$\cdot 10^{15}$ cm$^{-2}$. The mean value is approximately \linebreak 7$\cdot 10^{14}$ cm$^{-2}$, the median is about 5$\cdot 10^{14}$ cm$^{-2}$ (cf. Table \ref{IRDC_stat}). As the rotation temperature and the line width, also the column density of ammonia is an observable for the evolution of star-forming regions. Fig. \ref{Trot_density} shows that the column densities of our IRDCs are on average smaller than those of the HMPO sample (mean: 4$\cdot 10^{15}$ cm$^{-2}$, median: 3$\cdot 10^{15}$ cm$^{-2}$). This can be interpreted as evidence that the column densities of ammonia do change on spatial scales traced by single-dish telescopes in these early stages of massive star formation. The ammonia column densities of our sample are about one order of magnitude smaller than the ones of \citet{Wienen2012}, but agree with the ones of \citet{Harju1991,Harju1993} and \citet{Dunham2011}. Lower average column densities could be 
caused by our selection criteria based on absorption shadows. It appears that the extinction features trace the column density peaks less well than the ATLASGAL submm continuum observations. \\
		
		\subsection{Hydrogen Properties and Virial Parameter}\label{virial}
			\subsubsection{Virial Mass}\label{virial_mass}
				When we derived the virial mass M$_{\mbox{\tiny vir}}$, we only considered the balance between bulk kinetic and gravitational energy. Solving this for the mass and assuming a Gaussian velocity distribution, the virial mass can be calculated by 
				\begin{equation}
					\mbox{M}_{\mbox{\tiny vir}} = k_2 \, \mbox{r} \, \Delta \mbox{v}^2_1
					\label{M_vir_def}
				\end{equation}
				where r represents the radius of the source and $\Delta$v$_1$ its FWHM in ammonia (1,1)-inversion transition \citep{MacLaren1988}. \\ 
				\noindent $k_2$ is a constant depending on the assumed density profile (see Table \ref{tab_virial_coefficient}).
				\begin{table}[H]
					\caption{Virial Theorem Coefficient \citep{MacLaren1988}.}
					\begin{tabular}{cc}
						\hline
						Density Distribution & $k_2$ \\ \hline
						$\rho \sim r^{-1}$ & 190 \\
						$\rho \sim r^{-2}$ & 126 \\ \hline
					\end{tabular}
					\label{tab_virial_coefficient}
				\end{table}
				\noindent It is not clear whether the sources' density profiles follow a $r^{-1}$- or $r^{-2}$-power-law. Therefore, the masses M$_{vir}^{r^{-1}}$ and M$_{vir}^{r^{-2}}$ were calculated with both power-laws. They are presented in \linebreak Table \ref{tab02_initial}. For Table \ref{IRDC_stat} and the further calculations of the virial parameters (cf. Sect. \ref{virial_parameter}) we used the mean value $M_{\mbox{\tiny vir}} = \frac{1}{2} \left( M_{\mbox{\tiny vir}}^{r^{-1}}+M_{\mbox{\tiny vir}}^{r^{-2}} \right)$. \\
				\noindent The (beam averaged) virial masses range between 6 M$_\odot$ and 1100 M$_\odot$ with a mean value of approximately 325 M$_\odot$ and a median of 264 M$_\odot$. \\
			\subsubsection{Hydrogen Gas Mass}\label{virial_gas}
				For the calculation of the gas masses, we used the dust emission maps of the ATLASGAL project \citep{Schuller2009}, more precisely the sources' flux densities. For this part we made use of the Gildas GREG package \citep{gildas_greg}. Since the ATLASGAL maps have a HPBW of 19.2'' we need to smooth them. Otherwise we would not be able to compare them with our Effelsberg data (40'' HPBW). We did that by folding the ATLASGAL maps with a Gaussian of $\sqrt{\mbox{$(40'')^2\, - \,(19'')^2$}} = 35.1''$. Then, we measured the flux densities F$_\nu$ at the locations of the individual IRDC peaks. The errors are estimated to be 20\% of the measured flux densities. This is sufficient, because the gas masses' errors are dominated by the error of the dust absorption coefficient.
				The gas mass M$_{\mbox{\tiny gas}}$ is given by 
				\begin{equation}
					\mbox{M}_{\mbox{\tiny gas}} = \frac{\mbox{d}^2 \, \mbox{F}_\nu^{40''} \, R}{\mbox{B}_\nu(\mbox{T}_{\mbox{\tiny D}}) \, \kappa_\nu}
				\end{equation}
				where d is the source's distance to the sun, F$_\nu^{40''}$ the total flux density measured in dust emission at 870 $\mu$m \linebreak ($\nu$ = 345 GHz) within the smoothed beam of 40$''$. B$_\nu$(T$_{\mbox{\tiny D}}$) represents the Planck function for the dust temperature T$_{\mbox{\tiny D}}$ being approximately equal to the kinetic temperature T$_{\mbox{\tiny kin}}$ for temperatures below 50 K in the dense environment of IRDCs,  since gas and dust temperature are collisionally coupled at high densities in our IRDCs. Furthermore also the kinetic and the rotation temperatures are approximately equal below \linebreak 50 K. Thus, we inserted the rotation temperature T$_{\mbox{\tiny rot}}$ for the dust temperature. \\
				$R$ represents the gas-to-dust mass ratio and $\kappa_\nu$ the dust absorption coefficient. Following \citet{Schuller2009}, we assume $R$ to be 100 and $\kappa_{\mbox{\tiny OH}}$ = 1.85 cm$^2$ g$^{-1}$ \citep{Schuller2009,Ossenkopf1994}. But since these values are just estimated, they contain the biggest uncertainties, particularly the dust absorption coefficient. \\
				In the paper by \citet{Ossenkopf1994}, the dust absorption coefficient dependency on different conditions is described in more detail and the older data of \citet{MRN1977} are compared with newly derived values. \citeauthor{Ossenkopf1994} list the dust absorption coefficient in dependency of the wavelength and used gas density. The dust absorption coefficient of $\kappa_{\mbox{\tiny OH}}$ = 1.85 cm$^2$ g$^{-1}$ was interpolated from \linebreak Table 1, Col. 5 \citep{Ossenkopf1994,Schuller2009} at a wavelength of 870 $\mu$m. By interpolating the original data of \citeauthor{MRN1977} -- given in Table 1, Col. 2 \citep{Ossenkopf1994} -- we determined the dust absorption coefficient to be \linebreak $\kappa_{\mbox{\tiny MRN}}$ = (0.6 $\pm$ 0.1) cm$^2$ g$^{-1}$. Comparing this with $\kappa_{\mbox{\tiny OH}}$, this implies an error factor of about 3 in gas mass. Since the exact value is not known, the two models give us an estimate of the upper and lower limits of gas masses (cf. Table \ref{IRDC_stat}). We 
calculated the gas masses for all sample sources with both models, being labelled as M$_{\mbox{\tiny OH}}$ and M$_{\mbox{\tiny MRN}}$ (as being the corresponding virial parameters). \\
				Nevertheless, our sources with masses up to several thousand solar masses are on average more massive than the clumps of \citet{Harju1991,Harju1993}. This is not surprising, because our sample was selected for high-mass cores, while \citet{Harju1991,Harju1993} observed mainly low- to intermediate-mass cores.
			\subsubsection{Virial Parameter}\label{virial_parameter}
				The virial parameter for molecular clouds is defined by 
				\begin{equation}
					\alpha = a \, \frac{2 \mbox{T}}{|\mbox{W}|} = \frac{\mbox{M}_{\mbox{\tiny vir}}}{\mbox{M}_{\mbox{\tiny gas}}}
					\label{equ_alpha_def}
				\end{equation}
				\citep{Bertoldi1992} where the dimensionless parameter $a$ is the correction factor for sources having no uniform and spherical mass distribution. T and W represent the kinetic and gravitational energies. Their ratio is equivalent to the ratio of virial and gas mass. \\
				The virial parameter is a helpful tool, because it is an indicator for a source being in virial equilibrium or not. In a simplified picture, a source is in virial equilibrium if its virial parameter is equal to 1. If the virial parameter is bigger than 1, turbulence dominates over gravity and the source expands. Otherwise, if the virial parameter becomes smaller than 1, gravity is the dominating force and the cloud collapses. In reality, additional effects like gas pressure, external pressure and magnetic fields (being ignored in our calculations) influence a cloud's state of equilibrium. Even clouds with $\alpha \gg$ 1 can be stable, because these additional terms can change the balance \citep{Bertoldi1992}. \\
				\begin{figure}[t]
					\centering
					\subfloat[\citet{Ossenkopf1994}: $\kappa_{\mbox{\tiny OH}}$ = 1.85 cm$^2$ g$^{-1}$]{
						\includegraphics[width=0.5\textwidth]{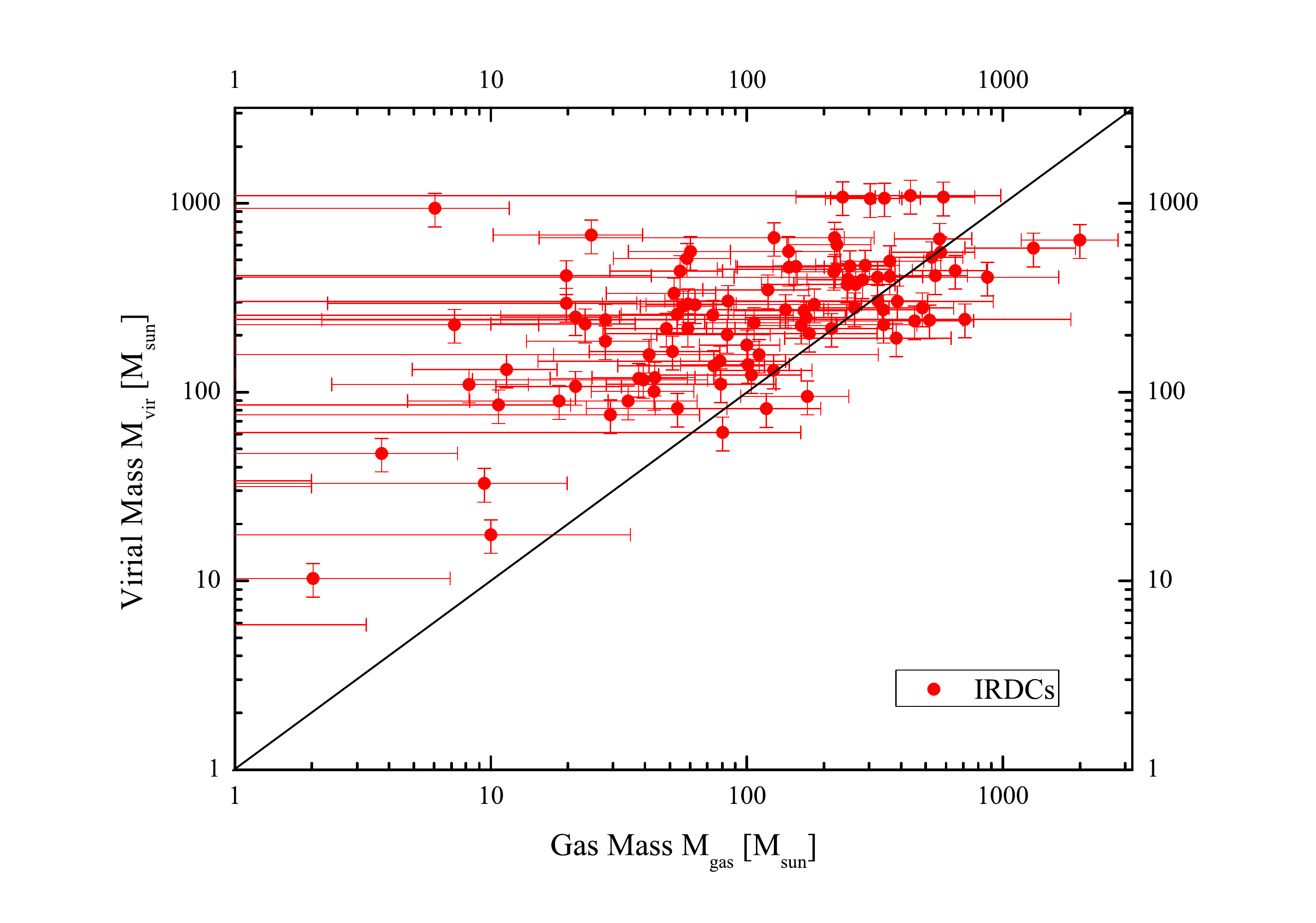}
						\label{Mvir_Moh}
					} \\
					\subfloat[\citet{MRN1977}: $\kappa_{\mbox{\tiny MRN}}$ = 0.6 cm$^2$ g$^{-1}$]{
						\includegraphics[width=0.5\textwidth]{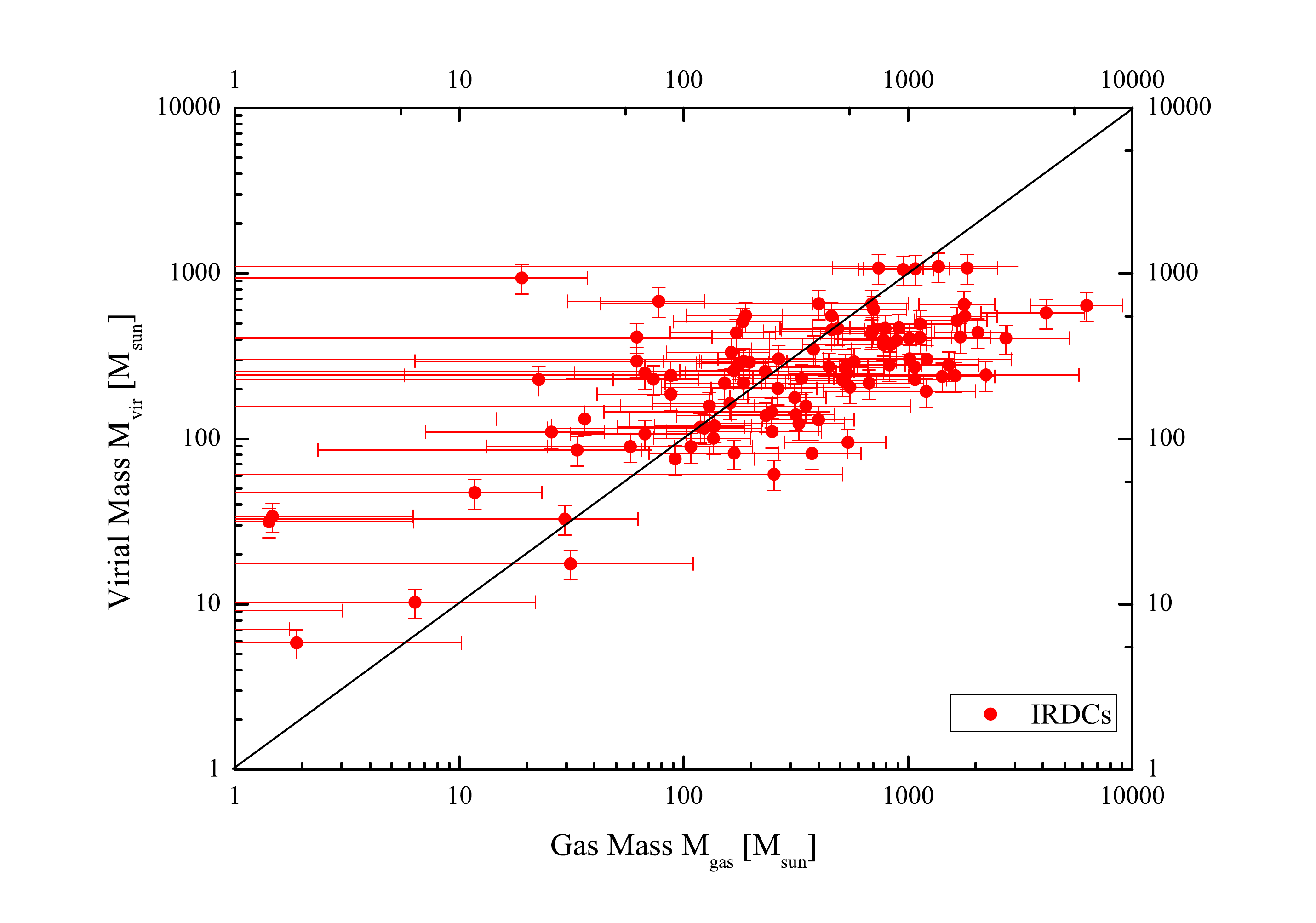}
						\label{Mvir_Mmrn}
					}
					\caption{Correlation plot between virial and gas mass using the model of \citet{Ossenkopf1994} and \citet{MRN1977}. The indicated black straight line represents the line of $\alpha$ being equal to 1.}
					\label{pic_vir_parameter}
				\end{figure}
				\noindent In Fig. \ref{pic_vir_parameter} the correlations between the virial and gas masses are plotted for each model. The value $\alpha$ = 1 is indicated by a straight black line. For the model of \citet{Ossenkopf1994} the majority of the IRDCs have virial parameters slightly above 1 (mean: 6, median: 2). In contrast, the majority of the sources' virial parameters in the model of \citet{MRN1977} are slightly below 1 (mean: 2, median 0.7). Because the models represent maximum and minimum estimates, the real virial parameters are approximately equal to 1 and the sources, thus, are close to virial equilibrium. This implies that the gas clumps can be stable, but may be at the verge of collapse and star formation. \\
				Our results match those of \citet{Dunham2011}, but not quite those of \citet{Wienen2012} and \citet{Harju1993}. \citet{Wienen2012} have mean virial parameters of 0.21 for FWHMs smaller than 4 km s$^{-1}$ (and 0.45 for FWHMs larger than \linebreak 4 km s$^{-1}$, but we do not have any source with such a large line width) by using the Ossenkopf \& Henning model and assuming the equipartition between magnetic and kinetic energy which reduces their value by a factor 0.5. Additionally, the difference is, to some degree, certainly due to their lower spectral resolution of 0.7 km s$^{-1}$ compared to our spectral resolution of \linebreak 0.38 km s$^{-1}$, although \citeauthor{Wienen2012} corrected for this in their calculation of virial masses. This reduces their used FWHM, and by that the estimate. \\
				In contrast, the virial parameters of \citet{Harju1993} \linebreak ($ = \frac{\mbox{2 E}_{\mbox{\tiny kin}}}{\mbox{E}_{\mbox{\tiny pot}}}$) are between 1.7 and 4 and, therefore, larger than ours. An explanation could be that their cores are more evolved than ours, indicating a more active phase of star formation. Since their sources can be associated with IRAS sources, this is not surprising. This may look like a contradiction to the small line widths (cf. Sect. \ref{width}). But considering the smaller solid angles and distances, \citet{Harju1991} observed also smaller absolute values of mass, kinetic and potential energies. Since the virial parameter is calculated by the ratio of kinetic and potential energies, the absolute measured order of magnitude does not really matter, but the relation between the energies. And this relation depends on the evolutionary stage of the objects.
			\subsubsection{Hydrogen Column Density and Ammonia Abundance}\label{ammonia_abu}
				With the ATLASGAL data, we were also able to calculate the column density of molecular hydrogen and the abundance of ammonia compared to hydrogen by using the formula
				\begin{equation}
					\mbox{N$_{H_2}^{19''}$} = \frac{\mbox{F}_\nu^{\mbox{\tiny 19}''} \, R}{\mbox{B}_\nu(\mbox{T}_{\mbox{\tiny D}}) \, \Omega \, \kappa_\nu \, \mu  \, \mbox{m}_{\mbox{\tiny H}}}
					\label{h2column}
				\end{equation}
				\citep{Schuller2009} where F$_\nu^{19''}$ is the total flux density measured in dust emission at 870 $\mu$m within the solid angle of $\Omega = 6.66 \cdot 10^{-9}$ sr (19$''$), $\mu$ the molecular weight of hydrogen and m$_{\mbox{\tiny H}}$ the weight of one hydrogen atom. For the dust absorption coefficient we adapted \linebreak $\kappa_\nu\,=$ 1.85 cm$^2$ g$^{-1}$ from \citet{Schuller2009}. The column densities of molecular hydrogen in our sample range between \linebreak 2$\cdot 10^{20}$ cm$^{-2}$ and 3$\cdot 10^{23}$ cm$^{-2}$, with an average of 4$\cdot 10^{22}$ cm$^{-2}$ and a median of 3$\cdot 10^{22}$ cm$^{-2}$ (cf. Table \ref{IRDC_stat}). These results match those of \citet{Wienen2012} and \citet{Dunham2011} \linebreak ($\sim 10^{22}$ cm$^{-2}$). \\
				Since the ammonia column densities have been calculated with data observed with a 40$''$ beam, we used the smoothed ATLASGAL data for deriving the ammonia abundances \linebreak $\chi_{NH_3}$ = N$_{NH_3}$ / N$_{H_2}^{40''}$. N$_{H_2}^{40''}$ is the hydrogen column density derived with Equ. (\ref{h2column}) by inserting the total flux density F$_\nu^{40''}$ within the solid angle $\Omega = 2.95 \cdot 10^{-8}$ sr (40$''$, smoothed data which have been also used for calculating the sample sources' gas masses, cf. Sect. \ref{virial_gas}). We find NH$_3$ abundances lie between 6$\cdot 10^{-9}$ and 5$\cdot 10^{-7}$ with a median of 2$\cdot 10^{-8}$. The average abundance is of the order $10^{-8}$ which agrees with the results of \citet{Wienen2012} and \citet{Dunham2011} ($\sim 5 \cdot 10^{-9} - 5 \cdot 10^{-7}$).
	\section{Summary and Conclusion}\label{summary}
		218 candidate IRDCs located between 13$^\circ$ and 50$^\circ$ in Galactic Longitude and $\pm$1.5$^\circ$ Galactic Latitude have been observed in the ammonia (1,1)- and (2,2)-inversion transitions with the Effelsberg 100m radio telescope. In 109 sources we found ammonia (1,1)-inversion lines and discussed the physical conditions within them. \\
		We found that there is a correlation between a source's position within the Galactic Plane and the detectability of ammonia. The smaller the absolute value of the source's Galactic Latitude, the greater the chance to find ammonia inversion lines. This makes sense, because ammonia is a high density tracer and the gas density is higher in the inner regions of the Galactic Plane than in the outer ones. To improve the detection rate, future observations should focus on areas of low Galactic Latitudes \linebreak ($|b| <$ 0.5$^\circ$). The dust emission maps of the ATLASGAL project confirm this statement, because the majority of compact sources detected by them have been found in this area \citep{Beuther2012}. Vice versa, almost all sources with high submm flux densities have counterparts in the IRDC catalogue \citep{Schuller2009}. Therefore, the detection rate could be increased by building up future IRDC samples on the base of the ATLASGAL data. We checked if our sources can be associated with Spitzer 24$\mu$m 
emission indicating the existence of Young Stellar Objects embedded within the IRDC candidates. We found the fraction of sample IRDCs having a counterpart in the Spitzer data to be 20.2\%, but were not able to find any correlation between the association and the temperatures measured with the ammonia observations. \\
		\noindent The observed line widths range between 0.5 and 2.5 km s$^{-1}$, some up to 4 km s$^{-1}$, and are much higher than expected for purely thermal broadening \linebreak ($\approx$ 0.2 km s$^{-1}$). Therefore, turbulence has to play an important role to stabilise IRDCs. Additionally, the FWHMs are significantly smaller than the ones of HMPOs, but higher than values measured in low-mass cores. We interpret this as additional evidence for IRDCs representing very early stages of high-mass star formation, although not all of our sample IRDCs will be able to actually form high-mass stars. The rotation temperatures (8 $\leq$ T$_{\mbox{\tiny rot}}$/K $\leq$ 30) are on average 15 K. Thus, IRDCs are cooler than HMPOs \linebreak ($\sim$ 22 K). The ammonia column densities of IRDCs are about an order of magnitude smaller than the ammonia column densities of HMPOs. \\
		Using ATLASGAL data, we computed the gas masses, column densities of molecular hydrogen and ammonia abundance relative to molecular hydrogen. We found molecular hydrogen column densities on the order of $10^{22}$ cm$^{-2}$ on average implying an average ammonia abundance on the order of $10^{-8}$. \\
		\noindent The virial masses derived for the cores within the IRDCs are between 100 and a few 1,000 M$_\odot$. The corresponding virial parameters are on the order of $\sim$1. Thus, the majority of observed IRDCs is close to virial equilibrium and hence consists of potential candidates for pre-protostellar regions of future star formation.

	\begin{acknowledgements}
		This research has made use of the catalogue of Infrared Dark Clouds and Cores based on MSX data and was built up by \citeauthor{Simon2006} as well as the dust emission maps of the ATLASGAL project. L. B. acknowledges support from CONICYT projects FONDAP 15010003 and Basal PFB-06.
	\end{acknowledgements}
\newpage
	\bibliographystyle{aa} 
	\bibliography{ref}

\appendix

\section{Derivation of used Formulas}\label{app_derive}
	\subsection{Rotation and Kinetic Temperature}\label{app_temp}
		As described in Sect. \ref{observation}, the hyperfine fit routine returns the best fit parameters of the brightness temperature, T$_{\mbox{\tiny mb}}$, optical depth, $\tau$, velocity of rest, $V_{\mbox{\tiny lsr}}$, and line width FWHM, $\Delta$v, (with errors). \\
		The brightness temperature T$_{\mbox{\tiny mb}}$ describes the brightness temperature of a source and depends on the beam this source is observed with by returning the mean value over this beam. It is needed to calculate the excitation temperature. \\			
		The excitation temperature T$_{\mbox{\tiny ex}}$ is no physical temperature, but describes the ratio between two population levels u(p) and l(ow) via
		\begin{equation}
			\frac{\mbox{n}_{\mbox{u}}}{\mbox{n}_{\mbox{\tiny l}}} = \frac{\mbox{g}_{\mbox{u}}}{\mbox{g}_{\mbox{\tiny l}}} \, \exp \left( - \frac{\Delta \mbox{E}}{k \mbox{T}_{\mbox{\tiny ex}}} \right)
			\label{equ_T_ex_def}
		\end{equation}
		where n$_{\mbox{\tiny i}}$ is the numbers of particles in the state i and g$_{\mbox{\tiny i}}$ its statistical weight. $\Delta$E represents the energy difference between the states and $k$ the Boltzmann constant. For the IRDCs temperatures above 10 K are expected. Thus, (by applying the Rayleigh-Jeans law) the brightness and excitation temperatures are simplified connected by
		\begin{equation}
			\mbox{T}_{\mbox{\tiny ex}} = \frac{\mbox{T}_{\mbox{\tiny mb}}}{\eta_{\mbox{f}} \, \left( \mbox{1} - \mbox{e}^{-\tau} \right)} + \mbox{T}_{\mbox{\tiny BG}}
			\label{equ_T_ex}
		\end{equation}
		with $\tau$ being the source's optical depth and T$_{\mbox{\tiny BG}}$ the background temperature (being about 2.7 K). $\eta_{\mbox{f}}$ represents the beam-filling factor. It describes the fraction of the antenna pattern being received from the source. In the case of extended and not clumpy sources, the beam-filling factor is equal to 1. We assume this case for the sample. The excitation temperatures are calculated for both transition lines. \\
		The rotation temperature T$_{\mbox{\tiny rot}}$ is defined similar to the excitation temperature by 
		\begin{equation}
			\frac{\mbox{n}^{\mbox{\tiny l}}_{\mbox{\tiny i}}}{\mbox{n}^{\mbox{\tiny l}}_{\mbox{\tiny j}}} = \frac{\mbox{g}_{\mbox{\tiny i}}}{\mbox{g}_{\mbox{\tiny j}}} \, \exp \left( - \frac{\Delta \mbox{E}}{k \mbox{T}^{\mbox{\tiny ij}}_{\mbox{\tiny rot}}} \right)
			\label{equ_T_rot_def}
		\end{equation}
		But in contrast to the excitation temperature, the rotation temperature does not describes the ratio between different levels being split by inversion, but between the levels of quantum numbers J and K (total angular momentum and its absolute projection along the z-axis). We are interested in the rotation temperatures of the metastable inversion levels with J $=$ K (non-metastable inversion levels: J $>$ K). These levels cannot be populated or depopulated by radiation. Therefore, their population numbers are high enough to emit measurable line intensities. Furthermore, metastable levels interact only via collisions \citep{Ho1983,Schilke} and are useful for the calculation of gas temperatures. \\
		The remaining problem is that we do not know the exact population numbers and have to approximate them by the column densities. For lines being observed in the same region the ratio of the population numbers should be equal to the ratio between the corresponding column densities depending on the ratio of the lines' optical depths: 
		\begin{equation}
			\frac{\tau^{\mbox{\tiny JK}}}{\tau^{\mbox{\tiny J'K'}}} \approx \frac{\mbox{K}^2}{\mbox{J}(\mbox{J}+\mbox{1})} \frac{\mbox{J'}(\mbox{J'}+\mbox{1})}{\mbox{K'}^2} \frac{\mbox{2J'}+\mbox{1}}{\mbox{2J}+\mbox{1}} \frac{\Delta \mbox{v}_1}{\Delta \mbox{v}_2} \frac{\mbox{N}_{\mbox{\tiny l}}(\mbox{J,K})}{\mbox{N}_{\mbox{\tiny l}}(\mbox{J',K'})} \frac{\nu_{\mbox{\tiny JK}}}{\nu_{\mbox{\tiny J'K'}}} \frac{\mbox{T}^{\mbox{\tiny J'K'}}_{\mbox{\tiny ex}}}{\mbox{T}^{\mbox{\tiny JK}}_{\mbox{\tiny ex}}}
			\label{equ_ratio}
		\end{equation}
		where $\tau^{\mbox{\tiny JK}}$ is the optical depth of the (J,K)-inversion line, $\Delta$v$_{\mbox{\tiny J}}$ its line width FWHM, N$_{\mbox{\tiny l}}$(J,K) its column density and T$^{\mbox{\tiny JK}}_{\mbox{\tiny ex}}$ its excitation temperature. Because only the (1,1)- and (2,2)-inversion lines have been studied, it is J = K = 2 and J' = K' = 1 in the following calculations. \\
		Inserting this into Equ. (\ref{equ_T_rot_def}), one gets:
		\begin{equation}
			\mbox{T}_{\mbox{\tiny rot}} = - \frac{\mbox{E}}{\mbox{x} \, \frac{\tau^{\mbox{\tiny JK}}}{\tau^{\mbox{\tiny J'K'}}} }
			\label{equ_T_rot_generell}
		\end{equation}
		where E $= \frac{h \, \nu_{\mbox{\tiny ij}}}{k} = $ 41.5 K \citep{Ho1983} and 
		\begin{equation}
			\mbox{x} = \frac{\mbox{K'}^2}{\mbox{J'(J'+1)}} \, \frac{\mbox{J(J+1)}}{\mbox{K}^2} \, \frac{\mbox{2J'}+\mbox{1}}{\mbox{2J}+\mbox{1}} \, \frac{\Delta \mbox{v}_2}{\Delta \mbox{v}_1} \, \frac{\nu_{\mbox{\tiny J'K'}}}{\nu_{\mbox{\tiny JK}}} \, \frac{\mbox{T}^{\mbox{\tiny JK}}_{\mbox{\tiny ex}}}{\mbox{T}^{\mbox{\tiny J'K'}}_{\mbox{\tiny ex}}}
			\label{equ_x_def}
		\end{equation}
		Following the instruction of \citet{Schilke} and \citet{Ho1983}, there are four cases one has to consider for calculating the rotation temperature. These cases differ in the optical depth of each inversion line:
		\begin{enumerate}
			\item both inversion lines are optical thick: \\
				\begin{equation}
					\mbox{T}_{\mbox{\tiny rot}} = - \frac{\mbox{E}}{\ln \left( \mbox{x} \frac{f_1}{f_2} \frac{\tau^{22}}{\tau^{11}} \right)}
					\label{equ_T_rot_1}
				\end{equation}
				where $\tau^{\mbox{\tiny ii}}$ is the optical depth of the (i,i)-inversion line and \linebreak $f_i$ is the relative intensity of the main hyperfine component. For the given transitions, there is $f_1 =$ 0.5 and $f_2 =$ 0.796 \citep{Ho1983}.
			\item only the (1,1)-inversion line is optical thick and the (2,2)-inversion line is optical thin: \\
				\begin{equation}
					\mbox{T}_{\mbox{\tiny rot}} = - \frac{\mbox{E}}{\ln \left[- \mbox{x} \frac{f_1}{\tau^{11} \, f_2} \ln \left( 1 - \frac{\tau^{22}}{\tau^{11}} \, \left( 1 - \mbox{e}^{-\tau^{11}} \right) \right) \right]}
					\label{equ_T_rot_2}
				\end{equation}
			\item both inversion lines are optical thin: \\
				\begin{equation}
					\mbox{T}_{\mbox{\tiny rot}} = - \frac{\mbox{E}}{\ln \left(- \mbox{x} \frac{f_1}{f_2} \frac{\mbox{T}_{\mbox{\tiny mb}}^{1}}{\mbox{T}_{\mbox{\tiny mb}}^{2}} \right)}
					\label{equ_T_rot_3}
				\end{equation}
				where $\mbox{T}_{\mbox{\tiny mb}}^{\mbox{\tiny i}}$ represents the brightness temperature of the (i,i)-inversion line.
			\item only the (1,1)-inversion line is detected, the (2,2)-inversion line is not: \\
				\begin{equation}
					\mbox{T}_{\mbox{\tiny rot}} = \frac{- \mbox{E}}{\ln \left[- \frac{\mbox{0.282}}{\tau^{11}} \ln \left( 1 - \frac{\mbox{T}_{\mbox{\tiny mb}}^{2}}{\mbox{T}_{\mbox{\tiny mb}}^{1}} \, \left( 1 - \mbox{e}^{-\tau^{11}} \right) \right) \right]}
					\label{equ_T_rot_4}
				\end{equation}
				In this case, the (2,2)-inversion line lies within the noise. To be able to continue with the calculations, $\mbox{T}_{\mbox{\tiny mb}}^{2}$ has been given the triple value of the average root-mean-square noise \linebreak (RMS = 0.0925 K). It is important to emphasise that in this case it is not possible to derive exact rotation temperatures, but only upper limit estimations! 
		\end{enumerate}
		Having the rotation temperature of ammonia, we were able to calculate the kinetic temperature of a source's gas by using the approximation of \citet{Tafalla2004}: \\
		\begin{equation}
			\mbox{T}_{\mbox{\tiny kin}} = \frac{\mbox{T}_{\mbox{\tiny rot}}}{ 1 - \frac{\mbox{T}_{\mbox{\tiny rot}}}{\mbox{E}} \, \ln \left[ 1 + 1.1 \, \exp \left( -  \frac{\mbox{15.7 K}}{\mbox{T}_{\mbox{\tiny rot}}} \right) \right] }
		\end{equation}
		This approximation has been derived with Monte Carlo models and gives an accuracy of 5\% in the range between 5 and 20 K. Because the majority of the sample IRDCs are within this range, their errors are set to this.
	\subsection{Column Density}\label{app_dens}
		As mentioned before in Sect. \ref{app_temp}, the column density is related to a source's optical depth \citep{Schilke}. If one neglects the background radiation and expresses the excitation temperature in terms of the optical depth, following \citet{Schilke}, the column density of the (1,1)-inversion level N$^{11}$ can by calculated by
		\begin{equation}
			\mbox{N}^{11} \approx 2 \mbox{T}_{\mbox{\tiny mb}} \, \frac{\tau}{1 - \mbox{e}^{-f_1 \, \tau^{11}}} \frac{\mbox{3} h}{\mbox{8} \pi^3} \frac{\sqrt{\pi}}{2 \sqrt{\ln (2)}} \frac{\mbox{J'(J'+1)}}{\mbox{K'}^2} \frac{\Delta \mbox{v}_1}{\mu^2} \frac{k}{h \nu_1}
			\label{equ_N_11}
		\end{equation}
		where $\mu =$ 1.476 Debye is the electric dipole moment and \linebreak $\nu_1 =$ 23694.496 MHz the laboratory frequency of the NH$_3$ (1,1)-inversion transition. To derive the total column density, N$_{NH_3}$, we assumed that the sources are in thermal equilibrium. In this case, the column density follows a Boltzmann distribution. Additionally, we took into account that Equ. (\ref{equ_N_11}) calculates the column density of para-NH$_3$ being the ammonia inversion levels with K $\neq$ 3n (n being an integer) where the hydrogen spins are not parallel. In contrast, the states with parallel hydrogen spins are called ortho-NH$_3$ (with K = 3n). The statistical weight g$_{\mbox{\tiny JK}}$ of ortho-NH$_3$ is twice the one of para-NH$_3$. Therefore, for calculating the total column density of ammonia, one has to take the triple of column density of para-NH$_3$ \linebreak \citep[cf.][]{Schilke}: 
		\begin{equation}
			\mbox{N}_{NH_3} = \frac{\mbox{3 N}^{11}}{\mbox{2(2J' + 1)}} \, \sum_{\mbox{1,1}} \mbox{g}_{11} \exp \left[ \left( \mbox{23.4 K} - \frac{h \nu_1}{k} \right) \, \frac{1}{\mbox{T}_{\mbox{\tiny rot}}} \right]
			\label{equ_N_tot}
		\end{equation}

\onecolumn
\thispagestyle{headings}
\newpage

	\section{Position of sample IRDCs within the Galactic Plane} \label{App_bigfig}
		\begin{figure}[H]
			\centering
			\includegraphics[angle=90,width=0.8\textwidth]{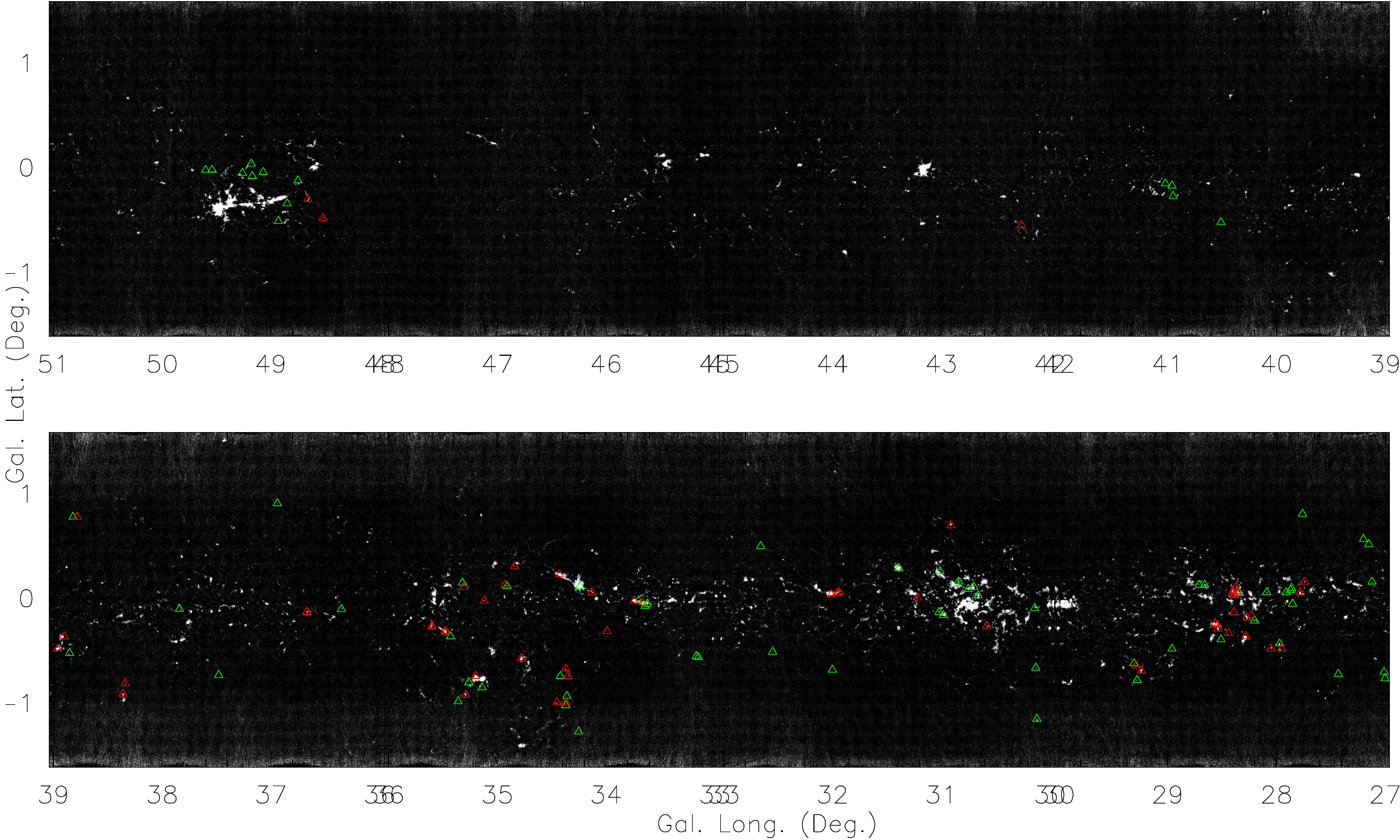}
			\caption{ATLASGAL dust emission map with indicated sample IRDCs in the range of \mbox{l = 27$^\circ$} to 51$^\circ$ \citep{Schuller2009}. The red triangles indicate the IRDCs with detected ammonia lines, the green triangles the ones without detected ammonia lines.}
			\label{App_fig01}
 		\end{figure}
		\newpage
 		\begin{figure}[H]
			\centering
			\includegraphics[angle=90,width=0.8\textwidth]{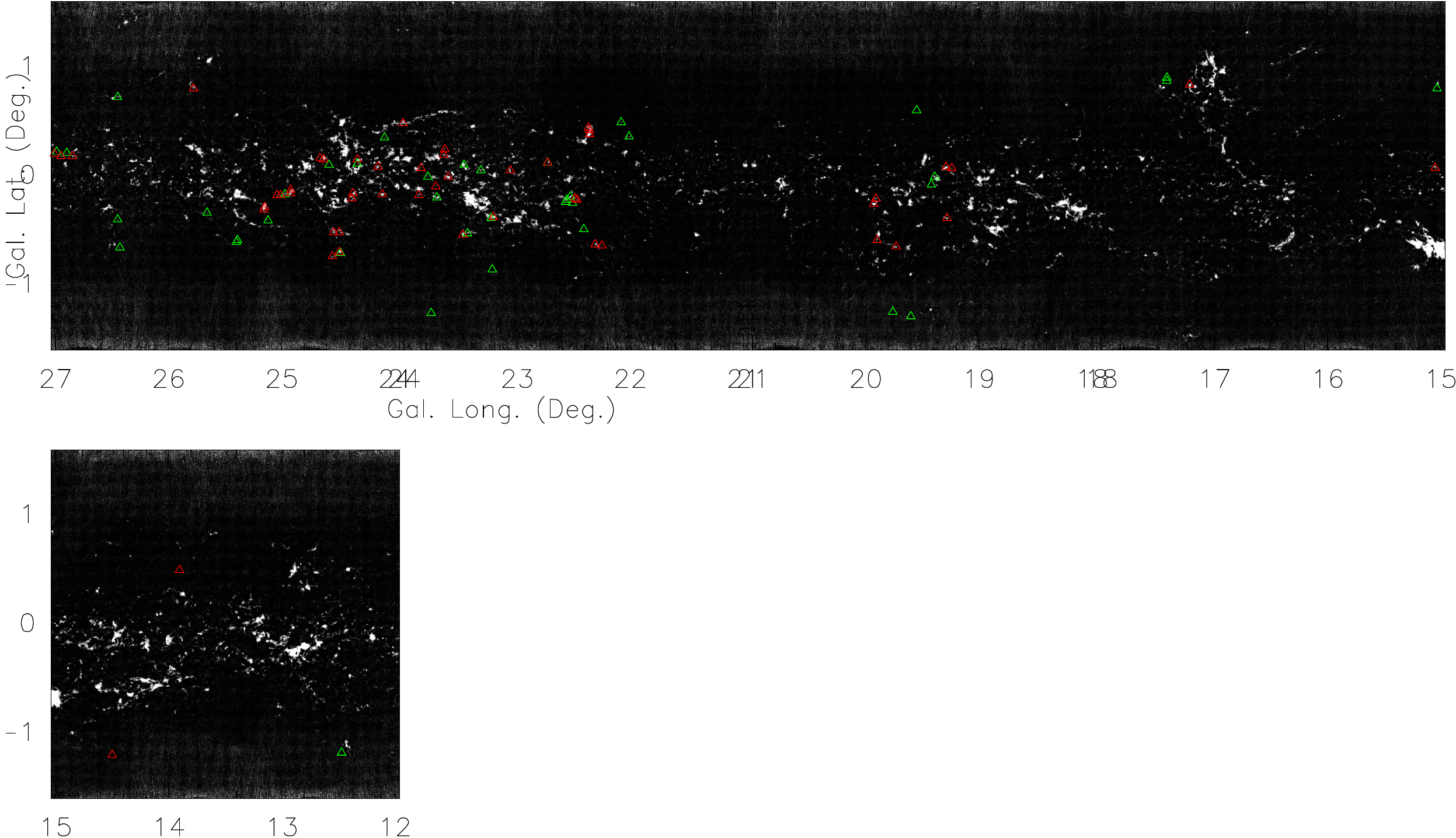}
			\caption{ATLASGAL dust emission map with indicated sample IRDCs in the range of \mbox{l = 12$^\circ$} to 27$^\circ$ \citep{Schuller2009}. The red triangles indicate the IRDCs with detected ammonia lines, the green triangles the ones without detected ammonia lines.}
			\label{App_fig02}
		\end{figure}

	\newpage

\section{Observational and calculated data of sample IRDCs}\label{app_data}
		\newcolumntype{d}{D{.}{.}{-1}}
		\renewcommand{\thefootnote}{\alph{footnote}}
\setlongtables
\begin{longtable}{l|dd|dddd|ddd|dddd}
\caption{Coordinates and fit parameters of detected IRDCs in NH$_3$. The values of the sources derived with or without detected (2,2)-inversion line are separated by a horizontal line. Note that latter give only upper limits. This table contains the sources' coordinates in Galactic longitude, $l$, and latitude, $b$, the brightness temperatures of both transition line, T$_{mb}^1$ and T$_{mb}^2$, the line widths of both lines, $\Delta v_1$ and $\Delta v_2$, the LSR velocity of the (1,1)-inversion line, V$_{\mbox{\tiny lsr}}^1$, and the optical depths of both lines, $\tau^{11}$ and $\tau^{22}$.} \\ \hline
Source & \multicolumn{1}{c}{$l$} & \multicolumn{1}{c}{$b$} & \multicolumn{1}{c}{T$_{mb}^1$} & \multicolumn{1}{c}{$\pm$} & \multicolumn{1}{c}{T$_{mb}^2$} & \multicolumn{1}{c}{$\pm$} & \multicolumn{1}{c}{$\Delta v_1$\footnotemark[5]} &  \multicolumn{1}{c}{$\Delta v_2$\footnotemark[5]} & \multicolumn{1}{c}{V$_{\mbox{\tiny lsr}}^1$\footnotemark[5]} & \multicolumn{1}{c}{$\tau^{11}$}  &   \multicolumn{1}{c}{$\pm$} & \multicolumn{1}{c}{$\tau^{22}$} & \multicolumn{1}{c}{$\pm$} \\ 
   & \multicolumn{2}{c}{deg} & \multicolumn{4}{c}{K} & \multicolumn{3}{c}{km s$^{-1}$} & \multicolumn{4}{c}{} \\ \hline
\endfirsthead \hline
\multicolumn{14}{c}{\small \slshape continued from previous page} \\ \hline
Source & \multicolumn{1}{c}{$l$} & \multicolumn{1}{c}{$b$} & \multicolumn{1}{c}{T$_{mb}^1$} & \multicolumn{1}{c}{$\pm$} & \multicolumn{1}{c}{T$_{mb}^2$} & \multicolumn{1}{c}{$\pm$} & \multicolumn{1}{c}{$\Delta v_1$\footnotemark[5]} &  \multicolumn{1}{c}{$\Delta v_2$\footnotemark[5]} & \multicolumn{1}{c}{V$_{\mbox{\tiny lsr}}^1$\footnotemark[5]} & \multicolumn{1}{c}{$\tau^{11}$}  &   \multicolumn{1}{c}{$\pm$} & \multicolumn{1}{c}{$\tau^{22}$} & \multicolumn{1}{c}{$\pm$} \\ 
   & \multicolumn{2}{c}{deg} & \multicolumn{4}{c}{K} & \multicolumn{3}{c}{km s$^{-1}$} & \multicolumn{4}{c}{} \\ \hline
\endhead
\hline \multicolumn{14}{c}{\small \slshape continued on next page}\\ \hline
\endfoot
\hline
\endlastfoot
G13.90+00.50A & \multicolumn{1}{c}{13.90} & \multicolumn{1}{c}{0.49} & \multicolumn{1}{c}{2.7} & \multicolumn{1}{c}{0.4} & \multicolumn{1}{c}{0.8} & \multicolumn{1}{c}{0.1} & \multicolumn{1}{c}{1.5} & \multicolumn{1}{c}{2.1} & \multicolumn{1}{c}{22.8} & \multicolumn{1}{c}{1.83} & \multicolumn{1}{c}{0.22} & \multicolumn{1}{c}{0.10} & \multicolumn{1}{c}{0.21} \\
G15.05+00.07A & \multicolumn{1}{c}{15.05} & \multicolumn{1}{c}{0.07} & \multicolumn{1}{c}{1.4} & \multicolumn{1}{c}{0.3} & \multicolumn{1}{c}{0.3} & \multicolumn{1}{c}{0.1} & \multicolumn{1}{c}{1.0} & \multicolumn{1}{c}{0.8} & \multicolumn{1}{c}{29.9} & \multicolumn{1}{c}{2.66} & \multicolumn{1}{c}{0.29} & \multicolumn{1}{c}{10.21} & \multicolumn{1}{c}{4.50} \\
G17.33+00.88A & \multicolumn{1}{c}{17.22} & \multicolumn{1}{c}{0.84} & \multicolumn{1}{c}{1.8} & \multicolumn{1}{c}{0.1} & \multicolumn{1}{c}{0.6} & \multicolumn{1}{c}{0.0} & \multicolumn{1}{c}{1.7} & \multicolumn{1}{c}{2.2} & \multicolumn{1}{c}{22.5} & \multicolumn{1}{c}{1.36} & \multicolumn{1}{c}{0.11} & \multicolumn{1}{c}{0.10} & \multicolumn{1}{c}{0.35} \\
G19.27+00.07A & \multicolumn{1}{c}{19.29} & \multicolumn{1}{c}{0.08} & \multicolumn{1}{c}{2.5} & \multicolumn{1}{c}{0.2} & \multicolumn{1}{c}{1.0} & \multicolumn{1}{c}{0.0} & \multicolumn{1}{c}{1.8} & \multicolumn{1}{c}{2.6} & \multicolumn{1}{c}{26.9} & \multicolumn{1}{c}{2.24} & \multicolumn{1}{c}{0.10} & \multicolumn{1}{c}{0.10} & \multicolumn{1}{c}{0.10} \\
G19.27+00.07B & \multicolumn{1}{c}{19.24} & \multicolumn{1}{c}{0.07} & \multicolumn{1}{c}{0.3} & \multicolumn{1}{c}{0.1} & \multicolumn{1}{c}{0.3} & \multicolumn{1}{c}{0.1} & \multicolumn{1}{c}{4.0} & \multicolumn{1}{c}{3.7} & \multicolumn{1}{c}{25.8} & \multicolumn{1}{c}{1.84} & \multicolumn{1}{c}{0.57} & \multicolumn{1}{c}{4.03} & \multicolumn{1}{c}{1.60} \\
\end{longtable}
\footnotetext[5]{The errors of $\Delta v_1$, $\Delta v_2$ and V$_{\mbox{\tiny lsr}}$ are on average smaller than 0.1 km s$^{-1}$.}

		\newpage
		\setlongtables
\begin{longtable}{l|ddd|d|dd|dddd|d}
\caption{Physical Conditions, distances and virial masses of detected IRDCs. The values of the sources derived with or without detected (2,2)-inversion line are separated by a horizontal line. Note that latter give only upper limits. This table contains the sources' ammonia rotation, $T_{rot}$, and kinetic temperatures, $T_{kin}$, column densities, N$_{NH_3}$, distances to the sun, d, and virial masses in dependency of the used density profile, M$_{vir}^{r^{-1}}$ for $\rho \sim r^{-1}$ and M$_{vir}^{r^{-2}}$ for $\rho \sim r^{-2}$. As well it contains a notice about the sources being associated with a Spitzer 24 $\mu$m emission sources. ``yes'' means that there is a Spitzer emission source within the Effelsberg beam at the location given in Table \ref{tab01_fit}; ``no'' means that there is not.} \\ \hline
Source & \multicolumn{1}{c}{$T_{rot}$} & \multicolumn{1}{c}{$\pm$} & \multicolumn{1}{c}{$T_{kin}$} & \multicolumn{1}{c}{N$_{NH_3}$} & \multicolumn{1}{c}{d} & \multicolumn{1}{c}{$\pm$} & \multicolumn{1}{c}{M$_{vir}^{r^{-1}}$} & \multicolumn{1}{c}{$\pm$} & \multicolumn{1}{c}{M$_{vir}^{r^{-2}}$} & \multicolumn{1}{c}{$\pm$} & \multicolumn{1}{c}{Spitzer(24$\mu$m)} \\ 
  & \multicolumn{3}{c}{K} & \multicolumn{1}{c}{10$^{15}$ cm$^{-2}$} & \multicolumn{2}{c}{kpc} & \multicolumn{4}{c}{M$_{\odot}$} &  \\ \hline
\endfirsthead \hline
\multicolumn{12}{c}{\small \slshape continued from previous page} \\ \hline
Source & \multicolumn{1}{c}{$T_{rot}$} & \multicolumn{1}{c}{$\pm$} & \multicolumn{1}{c}{$T_{kin}$} & \multicolumn{1}{c}{N$_{NH_3}$} & \multicolumn{1}{c}{d} & \multicolumn{1}{c}{$\pm$} & \multicolumn{1}{c}{M$_{vir}^{r^{-1}}$} & \multicolumn{1}{c}{$\pm$} & \multicolumn{1}{c}{M$_{vir}^{r^{-2}}$} & \multicolumn{1}{c}{$\pm$} & \multicolumn{1}{c}{Spitzer(24$\mu$m)} \\ 
  & \multicolumn{3}{c}{K} & \multicolumn{1}{c}{10$^{15}$ cm$^{-2}$} & \multicolumn{2}{c}{kpc} & \multicolumn{4}{c}{M$_{\odot}$} &  \\ \hline
\endhead
\hline \multicolumn{12}{c}{\small \slshape continued on next page}\\ \hline
\endfoot
\hline
\endlastfoot
G13.90+00.50A & \multicolumn{1}{c}{9.2} & \multicolumn{1}{c}{4.4} & \multicolumn{1}{c}{9.6} & \multicolumn{1}{c}{1.5} & \multicolumn{1}{c}{2.2} & \multicolumn{1}{c}{0.7} & \multicolumn{1}{c}{173} & \multicolumn{1}{c}{57} & \multicolumn{1}{c}{115} & \multicolumn{1}{c}{38} & \multicolumn{1}{c}{no} \\
G15.05+00.07A & \multicolumn{1}{c}{12.0} & \multicolumn{1}{c}{1.3} & \multicolumn{1}{c}{12.8} & \multicolumn{1}{c}{0.5} & \multicolumn{1}{c}{2.6} & \multicolumn{1}{c}{0.6} & \multicolumn{1}{c}{98} & \multicolumn{1}{c}{24} & \multicolumn{1}{c}{65} & \multicolumn{1}{c}{16} & \multicolumn{1}{c}{no} \\
G17.33+00.88A & \multicolumn{1}{c}{10.4} & \multicolumn{1}{c}{9.1} & \multicolumn{1}{c}{11.0} & \multicolumn{1}{c}{0.9} & \multicolumn{1}{c}{1.8} & \multicolumn{1}{c}{0.7} & \multicolumn{1}{c}{190} & \multicolumn{1}{c}{74} & \multicolumn{1}{c}{126} & \multicolumn{1}{c}{49} & \multicolumn{1}{c}{no} \\
G19.27+00.07A & \multicolumn{1}{c}{8.9} & \multicolumn{1}{c}{1.9} & \multicolumn{1}{c}{9.2} & \multicolumn{1}{c}{2.1} & \multicolumn{1}{c}{2.0} & \multicolumn{1}{c}{0.7} & \multicolumn{1}{c}{246} & \multicolumn{1}{c}{82} & \multicolumn{1}{c}{163} & \multicolumn{1}{c}{55} & \multicolumn{1}{c}{no} \\
G19.27+00.07B & \multicolumn{1}{c}{27.3} & \multicolumn{1}{c}{12.2} & \multicolumn{1}{c}{38.1} & \multicolumn{1}{c}{0.3} & \multicolumn{1}{c}{1.9} & \multicolumn{1}{c}{0.7} & \multicolumn{1}{c}{1129} & \multicolumn{1}{c}{406} & \multicolumn{1}{c}{749} & \multicolumn{1}{c}{269} & \multicolumn{1}{c}{no} \\
G19.27-00.39A & \multicolumn{1}{c}{14.7} & \multicolumn{1}{c}{2.0} & \multicolumn{1}{c}{16.3} & \multicolumn{1}{c}{0.3} & \multicolumn{1}{c}{3.5} & \multicolumn{1}{c}{0.5} & \multicolumn{1}{c}{157} & \multicolumn{1}{c}{23} & \multicolumn{1}{c}{104} & \multicolumn{1}{c}{15} & \multicolumn{1}{c}{yes} \\
\end{longtable}

		\newpage
		\setlongtables
\begin{longtable}{l|dd|dd|d|dd|dd|dd|dd}
\caption{Studies based on ATLASGAL data of detected IRDCs. The values of the sources derived with or without detected (2,2)-inversion line are separated by a horizontal line. Note that latter give only upper limits. The table contains the sources' flux density, F$_\nu^{19''}$ / F$_\nu^{40''}$, in dust emission at 870 $\mu$m within a 19$''$ / 40$''$ beam, hydrogen column density, N$_{H_2}^{19''}$ / N$_{H_2}^{40''}$, ammonia abundance within a 40$''$ beam by using $\kappa_{OH}$ = 1.85 cm$^2$ g$^{-1}$, $\chi_{NH_3}$, gas masses, M$_{\mbox{\tiny OH}}$ by using the Ossenkopf \& Henning model and M$_{\mbox{\tiny MRN}}$ by using the Mathis et al. model, and virial parameters, $\alpha_{\mbox{\tiny OH}}$ by using the Ossenkopf \& Henning model and $\alpha_{\mbox{\tiny MRN}}$ by using the Mathis et al. model.} \\ \hline
Source & \multicolumn{1}{c}{F$_\nu^{19''}$} & \multicolumn{1}{c}{F$_\nu^{40''}$} & \multicolumn{1}{c}{N$_{H_2}^{19''}$} & \multicolumn{1}{c}{N$_{H_2}^{40''}$} & \multicolumn{1}{c}{$\chi_{NH_3}$} & \multicolumn{1}{c}{M$_{\mbox{\tiny OH}}$} & \multicolumn{1}{c}{$\pm$} & \multicolumn{1}{c}{M$_{\mbox{\tiny MRN}}$} & \multicolumn{1}{c}{$\pm$} & \multicolumn{1}{c}{$\alpha_{\mbox{\tiny OH}}$}  & \multicolumn{1}{c}{$\pm$} &  \multicolumn{1}{c}{$\alpha_{\mbox{\tiny MRN}}$} & \multicolumn{1}{c}{$\pm$} \\ 
   & \multicolumn{2}{c}{Jy} & \multicolumn{2}{c}{10$^{22}$ cm$^{-2}$} & \multicolumn{1}{c}{10$^{-8}$} & \multicolumn{4}{c}{M$_\odot$} & \multicolumn{2}{c}{\tiny{$\kappa_{OH}$ = 1.85 cm$^2$ g$^{-1}$}} & \multicolumn{2}{c}{\tiny{$\kappa_{MRN}$ = 0.59 cm$^2$ g$^{-1}$}} \\ \hline
\endfirsthead \hline
\multicolumn{14}{c}{\small \slshape continued from previous page} \\ \hline
Source & \multicolumn{1}{c}{F$_\nu^{19''}$} & \multicolumn{1}{c}{F$_\nu^{40''}$} & \multicolumn{1}{c}{N$_{H_2}^{19''}$} & \multicolumn{1}{c}{N$_{H_2}^{40''}$} & \multicolumn{1}{c}{$\chi_{NH_3}$} & \multicolumn{1}{c}{M$_{\mbox{\tiny OH}}$} & \multicolumn{1}{c}{$\pm$} & \multicolumn{1}{c}{M$_{\mbox{\tiny MRN}}$} & \multicolumn{1}{c}{$\pm$} & \multicolumn{1}{c}{$\alpha_{\mbox{\tiny OH}}$}  & \multicolumn{1}{c}{$\pm$} &  \multicolumn{1}{c}{$\alpha_{\mbox{\tiny MRN}}$} & \multicolumn{1}{c}{$\pm$} \\ 
   & \multicolumn{2}{c}{Jy} & \multicolumn{2}{c}{10$^{22}$ cm$^{-2}$} & \multicolumn{1}{c}{10$^{-8}$} & \multicolumn{4}{c}{M$_\odot$} & \multicolumn{2}{c}{\tiny{$\kappa_{OH}$ = 1.85 cm$^2$ g$^{-1}$}} & \multicolumn{2}{c}{\tiny{$\kappa_{MRN}$ = 0.59 cm$^2$ g$^{-1}$}} \\ \hline
\endhead \hline
\multicolumn{14}{c}{\small \slshape continued from previous page} \\ \hline
\endfoot
\hline
\endlastfoot
G15.05+00.07A & \multicolumn{1}{c}{0.12} & \multicolumn{1}{c}{0.64} & \multicolumn{1}{c}{1.0} & \multicolumn{1}{c}{1.3} & \multicolumn{1}{c}{3.8} & \multicolumn{1}{c}{54} & \multicolumn{1}{c}{30} & \multicolumn{1}{c}{168} & \multicolumn{1}{c}{98} & \multicolumn{1}{c}{1.5} & \multicolumn{1}{c}{0.9} & \multicolumn{1}{c}{0.5} & \multicolumn{1}{c}{0.3} \\
G17.33+00.88A & \multicolumn{1}{c}{0.31} & \multicolumn{1}{c}{2.01} & \multicolumn{1}{c}{3.1} & \multicolumn{1}{c}{4.6} & \multicolumn{1}{c}{2.0} & \multicolumn{1}{c}{112} & \multicolumn{1}{c}{214} & \multicolumn{1}{c}{350} & \multicolumn{1}{c}{676} & \multicolumn{1}{c}{1.4} & \multicolumn{1}{c}{2.7} & \multicolumn{1}{c}{0.5} & \multicolumn{1}{c}{0.9} \\
G19.27-00.39A & \multicolumn{1}{c}{0.42} & \multicolumn{1}{c}{1.19} & \multicolumn{1}{c}{3.0} & \multicolumn{1}{c}{1.9} & \multicolumn{1}{c}{1.4} & \multicolumn{1}{c}{127} & \multicolumn{1}{c}{52} & \multicolumn{1}{c}{400} & \multicolumn{1}{c}{175} & \multicolumn{1}{c}{1.0} & \multicolumn{1}{c}{0.5} & \multicolumn{1}{c}{0.3} & \multicolumn{1}{c}{0.2} \\
G19.27+00.07A & \multicolumn{1}{c}{0.56} & \multicolumn{1}{c}{1.95} & \multicolumn{1}{c}{6.7} & \multicolumn{1}{c}{5.3} & \multicolumn{1}{c}{4.0} & \multicolumn{1}{c}{175} & \multicolumn{1}{c}{147} & \multicolumn{1}{c}{551} & \multicolumn{1}{c}{471} & \multicolumn{1}{c}{1.2} & \multicolumn{1}{c}{1.0} & \multicolumn{1}{c}{0.4} & \multicolumn{1}{c}{0.3} \\
G19.27+00.07B & \multicolumn{1}{c}{0.31} & \multicolumn{1}{c}{0.47} & \multicolumn{1}{c}{1.2} & \multicolumn{1}{c}{0.4} & \multicolumn{1}{c}{7.5} & \multicolumn{1}{c}{6} & \multicolumn{1}{c}{6} & \multicolumn{1}{c}{19} & \multicolumn{1}{c}{18} & \multicolumn{1}{c}{155.2} & \multicolumn{1}{c}{150.5} & \multicolumn{1}{c}{49.4} & \multicolumn{1}{c}{48.6} \\
\end{longtable}

		\newpage
		\setlongtables
\begin{longtable}{l|dd}
\caption{List of sources without detectable ammonia (1,1)-inversion transition. The table contains the names and coordinates in Galactic Longitude $l$ and Galactic Latitude $b$.} \\ \hline
Source & \multicolumn{1}{c}{Galactic Longitude $l$} & \multicolumn{1}{c}{Galactic Latitude $b$} \\ 
   & \multicolumn{2}{c}{deg} \\ \hline
\endfirsthead \hline
\multicolumn{3}{c}{\small \slshape continued from previous page} \\ \hline
Source & \multicolumn{1}{c}{Galactic Longitude $l$} & \multicolumn{1}{c}{Galactic Latitude $b$} \\ 
   & \multicolumn{2}{c}{deg} \\ \hline
\endhead
\hline \multicolumn{3}{c}{\small \slshape continued on next page}\\ \hline
\endfoot
\hline
\endlastfoot
G15.00+00.79A & \multicolumn{1}{c}{15.03} & \multicolumn{1}{c}{0.80} \\
G17.33+00.88B & \multicolumn{1}{c}{17.42} & \multicolumn{1}{c}{0.90} \\
G17.33+00.88C & \multicolumn{1}{c}{17.42} & \multicolumn{1}{c}{0.87} \\
G17.81+02.02A & \multicolumn{1}{c}{17.81} & \multicolumn{1}{c}{2.02} \\
G17.90+02.05A & \multicolumn{1}{c}{17.90} & \multicolumn{1}{c}{2.06} \\
G19.39-00.01A & \multicolumn{1}{c}{19.40} & \multicolumn{1}{c}{-0.01} \\
G19.42-00.08A & \multicolumn{1}{c}{19.40} & \multicolumn{1}{c}{-0.08} \\
G19.55+00.60A & \multicolumn{1}{c}{19.53} & \multicolumn{1}{c}{0.61} \\
\end{longtable}

\end{document}